\documentclass[prd,nofootinbib,showpacs,superscriptaddress, preprint]{revtex4-1}
\usepackage[T1]{fontenc}
\usepackage{amsmath,amssymb}
\usepackage{epsfig}
\usepackage{graphicx}
\usepackage[usenames,dvipsnames]{color}
\usepackage{slashed}
\usepackage[colorlinks,citecolor=blue]{hyperref}
\usepackage{color}

\newcommand{\Tr}[1]{\text{Tr}\big[#1\big]}
\newcommand{\hc}{\ensuremath{\text{h.c.}}}
\def\be{\begin{equation}}
\def\ee{\end{equation}}
\def\bsp#1\esp{\begin{split}#1\end{split}}

\def\bpm{\begin{pmatrix}}
	\def\epm{\end{pmatrix}}

\begin{document}
\title{Left Right Symmetric Models with a Mixture of keV-TeV Dark Matter}

\author{Debasish Borah}
\email{dborah@iitg.ac.in}
\affiliation{Department of Physics, Indian Institute of Technology Guwahati, Assam 781039, India}
\author{Arnab Dasgupta}
\email{arnabdasgupta28@gmail.com}
\affiliation{Theoretical Physics Division, Physical Research Laboratory, Navrangpura, Ahmedabad, Gujarat 380009, India}
\affiliation{School of Liberal Arts, Seoul-Tech, Seoul 139-743, Korea}
\begin{abstract}
We discuss the possibility of realising a multi-component dark matter scenario with widely separated dark matter masses: one having keV scale mass and the other with GeV-TeV scale mass, within the framework of left right symmetric models. Due to gauge interactions, both the dark matter candidates are produced thermally in the early Universe but overproducing the keV mass candidate. We consider one of the right handed neutrinos to be decaying at late epochs, just before the big bang nucleosynthesis, in order to dilute the thermally overproduced keV dark matter. We constrain the parameter space from the requirement of producing sub-dominant keV-TeV dark matter, satisfying indirect detection constraints from gamma ray searches and producing the tantalising 3.55 keV monochromatic X-ray line, reported by several groups to be present in galaxy and galaxy cluster data, from the decay of a 7.1 keV dark matter on cosmological scales. We find that these requirements can keep the right sector gauge boson masses around a few TeV while requiring some of the right handed neutrinos in the sub GeV regime.

\end{abstract}
\maketitle

\section{Introduction}
In the last few decades, there have been significant amount of hints and evidences suggesting the presence of non-luminous and non-baryonic form of matter (popularly known as dark matter (DM)) in the present Universe. Starting from the galaxy cluster observations by Fritz Zwicky \cite{Zwicky:1933gu} back in 1933, observations of galaxy rotation curves in 1970's \cite{Rubin:1970zza}, the more recent observation of the bullet cluster \cite{Clowe:2006eq} to the latest cosmology data provided by the Planck satellite \cite{Ade:2015xua}, the astrophysics, cosmology as well as the particle physics community have a come a long way. The latest data from the Planck mission suggest that around $26\%$ of the present Universe's energy density is in the form of dark matter. In terms of density parameter and $h = \text{(Hubble Parameter)}/(100 \;\text{km} \text{s}^{-1} \text{Mpc}^{-1})$, the present dark matter abundance is conventionally reported as \cite{Ade:2015xua}
\begin{equation}
\Omega_{\text{DM}} h^2 = 0.1198 \pm 0.0015.
\label{dm_relic}
\end{equation}
Since none of the particles in the standard model (SM) of particle physics can serve as a DM candidate, it has lead to a plethora of DM models within several beyond standard model (BSM) frameworks. Although the SM neutrinos satisfy some of these criteria, yet they remain relativistic at the epoch of freeze-out as well as matter radiation equality, giving rise to Hot Dark Matter (HDM) which is ruled out by both astrophysics and cosmology observations. Among different BSM frameworks for viable DM candidates, the most popular or the most widely studied scenario perhaps, is the so called weakly interacting massive particle (WIMP) paradigm. In this framework, a dark matter candidate typically with electroweak scale mass and interaction rate similar to electroweak interactions can give rise to the correct dark matter relic abundance, a remarkable coincidence often referred to as the \textit{WIMP Miracle}. Such interactions kept the WIMP DM in thermal equilibrium in the early Universe and eventually its number density gets frozen out when the rate of expansion of the Universe takes over the interaction rates. Such DM candidates typically remain non-relativistic at the epoch of freeze-out as well as matter radiation equality and belong to the category of Cold Dark Matter (CDM). For a recent review of DM models based on WIMP paradigm, please see \cite{Arcadi:2017kky}.

The sizeable interactions of WIMP DM with other SM particles can not only generate its relic abundance after thermal freeze-out naturally, but also enhances the testability of it as such a DM particle can scatter off nuclei kept in a typical detector. However, till date no such DM-nucleon scattering has been observed in any of the experiments. The most recent dark matter direct detection experiments like LUX, PandaX-II and Xenon1T have also reported their null results \cite{Akerib:2016vxi, Tan:2016zwf, panda2017, Aprile:2017iyp}. Similar null results have been also reported by other direct search experiments like the large hadron collider (LHC) giving upper limits on DM interactions with the SM particles. A recent summary of collider searches for DM can be found in \cite{Kahlhoefer:2017dnp}. Although such null results could indicate a very constrained region of WIMP parameter space, they have also motivated the particle physics community to look for beyond the thermal WIMP paradigm. One interesting scenario is the kind of DM which remains mildly relativistic at the epoch of matter radiation equality, keeping it at intermediate stage between HDM and CDM and referred to as Warm Dark Matter (WDM). They typically have mass in the keV range, in contrast to HDM with sub-eV mass and CDM with GeV-TeV scale mass. For a recent review on WDM, one can refer to \cite{Adhikari:2016bei}. Such a scenario is particularly interesting as it can address several challenges like the missing satellite problem, too big to fail problem related to small scale structure formation, that arise in a CDM framework. For a recent review on these small scale challenges, please refer to \cite{Bullock:2017xww}. The classification of Hot, Warm and Cold DM is primarily done on the basis of their free streaming lengths which is roughly the distance for which the DM particles can freely propagate. For detailed calculation of free streaming lengths, please refer to \cite{Boyarsky:2008xj, Merle:2013wta}. Typically, the free streaming length $\lambda_{\text{FS}} =0.1$ Mpc, about the size of a dwarf galaxy, acts as a boundary line between HDM ($\lambda_{\text{FS}} >0.1$ Mpc) and WDM ($\lambda_{\text{FS}} <0.1$ Mpc). For CDM, on the other hand, the free streaming lengths are considerably smaller than this value. Therefore, CDM structures keep forming till scales as small as the solar system which gives rise to disagreement with observations at small scales \cite{Bullock:2017xww}. HDM, on the other hand, erases all small scale structure due to its large free streaming length, disfavouring the bottom up approach of structure formation. WDM can therefore act as a balance between the already ruled out HDM possibility and the CDM paradigm having issues with small scale structures.

Apart from these motivations, there are motivations from indirect detection experiments as well. There have been many efforts to look for indirect dark matter signatures at different experiments with the hope that even though dark matter may not scatter off nuclei significantly as indicated by the null results at direct detection experiments, but they may decay or annihilate into the standard model particles on cosmological scales and leave some indirect signatures. Interestingly, there have been some recent observations at some of these indirect detection experiments which could have possible dark matter origins. One promising indirect signature of dark matter was reported by two independent analysis \cite{Bulbul:2014sua} and \cite{Boyarsky:2014jta} of the data collected by the XMM-Newton X-ray telescope. Their analysis hinted towards the existence of a monochromatic X-ray line with energy 3.55 keV in the spectrum of 73 galaxy clusters. The analysis \cite{Bulbul:2014sua} also claimed the presence of the same line in the Chandra observations of the Perseus cluster. Later on, the same line was also found in the Milky Way by analysing the XMM-Newton data \cite{Boyarsky:2014ska}. Although the analysis of the preliminary data collected by the Hitomi satellite (before its unfortunate crash) do not confirm such a monochromatic line \cite{Aharonian:2016gzq}, one still needs to wait for a more sensitive observation with future experiments to have a final word on it. Interestingly, the authors of \cite{Conlon:2016lxl} considered a specific dark matter model to show consistency among Hitomi, XMM-Newton and Chandra observations. More recently, the authors of \cite{Cappelluti:2017ywp} have reported a $3\sigma$ detection of a $3.55$ keV emission line in the spectrum of the Cosmic X-ray background using Chandra observations towards the COSMOS Legacy and CDFS survey fields. Such a signal, if confirmed in future experiments, can be naturally explained by a keV scale sterile neutrino WDM that has mixing with the SM neutrinos of the order $\approx 10^{-11}-10^{-10}$ \cite{Bulbul:2014sua, Boyarsky:2014jta}. Different possible keV DM scenarios that can give rise to such an X-ray line were also discussed, for example \cite{Arcadi:2014dca}. One can also generate such a signal in typical WIMP DM models if there are two quasi-degenerate DM candidates having mass splitting of 3.55 keV, allowing the heavier one to decay into the lighter one and a photon. One such work can be found in \cite{Borah:2015rla}. The alternative possibility of keV dark matter annihilation into monochromatic photons was also discussed very recently by the authors of \cite{Brdar:2017wgy}. Although such keV scale WDM can not be detected in typical direct detection experiments like LUX, PandaX-II and Xenon1T, there have been some interesting proposals for direct detection of such light DM candidates. For example, one may refer to this recent article on direct detection prospects of sub-MeV DM \cite{Hochberg:2017wce}. There have also been serious attempts to look for keV sterile neutrino signatures in electron capture as well as beta decay spectra \cite{deVega:2011xh, Moreno:2016hrs}.

Instead of completely giving up on the CDM framework due to the negative results at dark matter direct detection as well as collider experiments, here we consider an exotic scenario where the dark sector consists of both cold and warm components. Such a mixed dark matter model can be very interesting from astrophysical structure point of view, as it may provide a way to solve the small scale structure problem \cite{Harada:2014lma}. However, there have not been many works regarding mixed DM scenarios with different thermal histories. In \cite{DuttaBanik:2016jzv}, the authors considered a mixed DM scenario where one candidate is of WIMP type whereas the other has a non-thermal origin due to its feeble interactions. Such a scenario has more optimistic detection prospects as it can be probed at both types of experiments: sensitive to sub-MeV as well as electroweak scale DM. From model building perspective, it may however be challenging to come up with realistic models that can account for such multi-component DM scenario. Since, both the DM components should be long lived or stable on cosmological scales, one may require exotic or non-minimal symmetries to guarantee that. For example a discrete unbroken $Z_2 \times Z_2$ symmetry can stabilise two DM components. Apart from the stability issue, another important aspect such models should have is a consistent production mechanism of DM. The CDM component, if belongs to the WIMP type DM, can be thermally produced in the early Universe. On the other hand, the production mechanism of WDM depends on the particular realisation and can have either thermal or non-thermal origin. For a summary of these production mechanisms, one can refer to this recent review article \cite{Adhikari:2016bei}. Here we consider a particle physics framework which naturally takes care of both the stability and production issue of the DM components. This is based on the left right symmetric model (LRSM) framework where the SM gauge symmetry is extended to $SU(3)_c \times SU(2)_L \times SU(2)_R \times U(1)_{B-L}$ which has been studied very extensively in the last few decades. Apart from the usual motivations for LRSM, here we have more motivations from DM point of view. Such an enlarged gauge symmetry can not only guarantee the stability of DM but also can ensure their productions in the early Universe by virtue of their gauge interactions. We consider CDM belonging to both left and right sectors of LRSM but keep the WDM part to the right sector only. This choice is particularly made in order to avoid severe electroweak precision constraints on introducing new keV scale particles having electroweak gauge interactions. We check how the requirement of generating a specific percentage of dark matter in terms of WDM affects the CDM parameter space and vice versa. The thermal relic of WDM is found to be more than the required DM abundance, requiring entropy dilution at later epochs \cite{Scherrer:1984fd} \footnote{Another interesting way to bring down the over-abundance incorporating a non-standard cosmological phase was proposed recently in \cite{Biswas:2018iny}.}. Since CDM freezes out at temperatures of GeV scale, the thermal relic abundance of CDM also gets affected by the late time entropy dilution required to bring the overproduced WDM within Planck limits. We therefore, look at that part of the CDM parameter space which overproduce it, so that after the entropy dilution, it can give rise to some sizeable fraction of total dark matter density. Since overproduction of WIMP typically involves smaller annihilation cross section, such CDM scenarios can easily evade strict constraints from indirect and direct detection experiments. Also, such a scenario will allow heavy mass region of any CDM scenarios where relic density often gets overproduced due to the unitarity bound on DM annihilations \cite{Nussinov:2014qva}, discussed recently in the context of PeV scale left-right DM in \cite{Borah:2017xgm}. Therefore, our analysis gives rise to completely new region of parameter space compared to single component DM. We also discuss how such a scenario can have interesting indirect detection prospects in both gamma-ray and X-ray experiments.

This article is organised as follows. In section \ref{sec1}, we briefly discuss the minimal LRSM and then briefly discuss the possibility of mixed dark matter in LRSM in section \ref{sec2}.  We discuss our results in section \ref{sec6} and finally conclude in section \ref{sec7}.

\section{Minimal Left-Right Symmetric Model (MLRSM)}
\label{sec1}
Left-Right Symmetric Model \cite{Pati:1974yy, Mohapatra:1974gc, Senjanovic:1975rk, Mohapatra:1980qe, Gunion:1989in, Deshpande:1990ip} is one of the well studied and well motivated BSM frameworks where the gauge symmetry of the electroweak theory is extended to $SU(3)_c \times SU(2)_L \times SU(2)_R \times U(1)_{B-L}$. The right handed fermions are doublets under $SU(2)_R$ similar to the way left handed fermions transform as doublets under $SU(2)_L$, in order to treat them on equal footing. The requirement of an anomaly free $U(1)_{B-L}$ makes the presence of three right handed neutrinos a necessity rather than a choice. This is in contrast with the type I seesaw models where three right handed singlet neutrinos are added by hand in order to generate light neutrino masses through seesaw mechanism. In MLRSM, to allow Dirac Yukawa couplings between $SU(2)_{L,R}$ doublet fermions, the Higgs field has to transform as a bidoublet under $SU(2)_{L,R}$ gauge symmetry. In order to break the gauge symmetry of the model to that of the SM spontaneously, scalar triplet fields with non-zero $U(1)_{B-L}$ charges are introduced, which also give Majorana masses to the left and right handed neutrinos.

The fermion content of the MLRSM is given by
\begin{equation}
Q_L=
\left(\begin{array}{c}
\ u_L \\
\ d_L
\end{array}\right)
\sim (3,2,1,\frac{1}{3}),\hspace*{0.8cm}
Q_R=
\left(\begin{array}{c}
\ u_R \\
\ d_R
\end{array}\right)
\sim (3^*,1,2,\frac{1}{3}),\nonumber 
\end{equation}
\begin{equation}
\ell_L =
\left(\begin{array}{c}
\ \nu_L \\
\ e_L
\end{array}\right)
\sim (1,2,1,-1), \quad
\ell_R=
\left(\begin{array}{c}
\ \nu_R \\
\ e_R
\end{array}\right)
\sim (1,1,2,-1) \nonumber.
\end{equation}
Similarly, the scalar content of the MLRSM is
\begin{equation}
\Phi=
\left(\begin{array}{cc}
\ \phi^0_{11} & \phi^+_{11} \\
\ \phi^-_{12} & \phi^0_{12}
\end{array}\right)
\sim (1,2,2,0)
\nonumber 
\end{equation}
\begin{equation}
\Delta_L =
\left(\begin{array}{cc}
\ \delta^+_L/\surd 2 & \delta^{++}_L \\
\ \delta^0_L & -\delta^+_L/\surd 2
\end{array}\right)
\sim (1,3,1,2), \hspace*{0.2cm}
\Delta_R =
\left(\begin{array}{cc}
\ \delta^+_R/\surd 2 & \delta^{++}_R \\
\ \delta^0_R & -\delta^+_R/\surd 2
\end{array}\right)
\sim (1,1,3,2) \nonumber
\end{equation}
where the numbers in brackets denote the transformations of the fields under the gauge group $SU(3)_c\times SU(2)_L\times SU(2)_R \times U(1)_{B-L}$ of the model. During the spontaneous symmetry breaking of MLRSM gauge group down to the SM gauge group, the neutral component of the Higgs triplet $\Delta_R$ acquires a non-zero vacuum expectation value (vev) after which the neutral components of Higgs bidoublet $\Phi$ acquire non-zero vev's to break the SM gauge symmetry into the $U(1)$ of electromagnetism. This symmetry breaking chain can be denoted as:
$$ SU(2)_L \times SU(2)_R \times U(1)_{B-L} \quad \underrightarrow{\langle
\Delta_R \rangle} \quad SU(2)_L\times U(1)_Y \quad \underrightarrow{\langle \Phi \rangle} \quad U(1)_{em}$$
Denoting the vev of the two neutral components of the bidoublet as $k_1, k_2$ and that of triplets $\Delta_{L, R}$ as $v_{L, R}$ and considering $g_L=g_R$,  $k_2 \sim v_L \approx 0$ and $v_R \gg k_1$, the approximate expressions for gauge boson masses after symmetry breaking can be written as 
$$ M^2_{W_L} = \frac{g^2}{4} k^2_1, \;\;\; M^2_{W_R} = \frac{g^2}{2}v^2_R $$
$$ M^2_{Z_L} =  \frac{g^2 k^2_1}{4\cos^2{\theta_w}} \left ( 1-\frac{\cos^2{2\theta_w}}{2\cos^4{\theta_w}}\frac{k^2_1}{v^2_R} \right), \;\;\; M^2_{Z_R} = \frac{g^2 v^2_R \cos^2{\theta_w}}{\cos{2\theta_w}} $$
where $\theta_w$ is the Weinberg angle. If we consider tiny but non-zero $k_2$, it gives rise to a left-right mixing between $W_L-W_R$ given by 
\begin{equation}
\tan{\theta_{LR}} = -\frac{2k_1 k_2}{v^2_R}
\end{equation}
Even if we switch off the vev $k_2$, then also there can be non-zero $W_L-W_R$ mixing, generated at one loop level. This can be calculated as \cite{Borah:2016hqn, Borah:2017leo}
\begin{align}
\sin{2\theta_{LR}} &=\frac{2W_{LR}}{\sqrt{\left(M^2_{W_R}-M^2_{W_L}\right)^2 + 4W^2_{LR}}} \nonumber \\ 
W_{LR} &= \frac{4\pi \alpha}{\sin^2 \theta_{W}}\sum_{u,d}m_u m_d V_{u,d}V^*_{u,d}f(x_{u,d}); \quad x_{i,j} = \frac{m^2_i}{m^2_j}\nonumber \\
f(x_{i,j}) &= \frac{1}{16\pi^2}\left[\frac{x_{i,j}\ln (x_{i,j}) + 1 - x_{i,j}}{1-x_{i,j}}+\ln \left(\frac{\mu^2}{m^2_{j}}\right)\right] 
\label{WLWRmixing}
\end{align}

The relevant Yukawa couplings for fermion masses can be written as 
\begin{eqnarray}
{\cal L}^{II}_\nu &=& y_{ij} \bar{\ell}_{iL} \Phi \ell_{jR}+ y^\prime_{ij} \bar{\ell}_{iL}
\tilde{\Phi} \ell_{jR} +Y_{ij} \bar{q}_{iL} \Phi q_{jR}+ Y^\prime_{ij} \bar{q}_{iL}
\tilde{\Phi} q_{jR} +\text{h.c.}
\nonumber \\
&+& f_{ij}\ \left(\ell_{iR}^T \ C \ i \sigma_2 \Delta_R \ell_{jR}+
(R \leftrightarrow L)\right)+\text{h.c.}
\label{treeY}
\end{eqnarray}
where $\tilde{\Phi} = \tau_2 \Phi^* \tau_2$. The scalar potential of the model is shown in appendix \ref{appen1}. In the above Yukawa Lagrangian, the indices $i, j = 1, 2, 3$ correspond to the three generations of fermions. The Majorana Yukawa couplings $f$ is same for both left and right handed neutrinos
because of the in built left-right symmetry $(f_L = f_R)$. These couplings $f$ give rise to the Majorana mass terms of both left handed and right handed neutrinos after the triplet Higgs fields $\Delta_{L,R}$ acquire non-zero vev. Although it is the $\Delta_R$ field which gets a vev at high scale breaking the left-right symmetry, the subsequent electroweak symmetry breaking induces a non-zero vev to the left handed counterpart. The induced vev for the left-handed triplet $v_{L}$ can be shown for generic LRSM to be
$$v_{L}=\gamma \frac{M^{2}_{W_L}}{v_{R}}$$
with $M_{W_L}\sim 80.4$ GeV being the weak boson mass such that 
$$ |v_{L}|<<M_{W_L}<<|v_{R}| $$ 
In general $\gamma$ is a function of various couplings in the scalar potential of generic LRSM. Using the results from Deshpande et al., \cite{Deshpande:1990ip}, $\gamma$ is given by
\begin{equation}
\gamma = \frac{\beta_2 k^2_1+\beta_1 k_1 k_2 + \beta_3  k^2_2}{(2\rho_1-\rho_3)(k^2_1+k^2_2)}
\label{eq:gammaLR}
\end{equation}
where $\beta, \rho$ are dimensionless parameters of the scalar potential. Without any fine tuning $\gamma$ is expected to be of the order unity ($\gamma\sim 1$). However, for TeV scale type I+II seesaw, $\gamma$ has to be fine-tuned as we discuss later. 

The $6\times6$ light+heavy neutrino
	mass matrix is then given, in the $(\nu_L, \nu_R)$ gauge eigenbasis, by
	\be
	M = \bpm \sqrt{2} f_L v_L & M_D \\ M^T_D & M_R \epm \ =\bpm M_{LL} & M_D \\ M^T_D & M_{RR} \epm \ 
	\ee
	Assuming $M_{LL}\ll M_D\ll M_R$, the light neutrino mass after symmetry breaking is generated within a type I+II seesaw as,
	\begin{equation}\label{eq9}
	\rm M_\nu= {M_\nu}^{I}+{M_\nu}^{II}
	\end{equation}
	\begin{equation}\label{eq6}
	M_\nu=M_{LL}-M_D{M_{RR}}^{-1}{M_D}^T
	=\sqrt{2}v_Lf_L-\frac{v^2_{\rm SM}}{\sqrt{2}v_R}h_D{f_R}^{-1}{h_D}^T,
	\end{equation}
	\begin{equation}\label{eq7}
	M_D=\frac{1}{\sqrt{2}}(k_1 y+k_2 y'), M_{LL}=\sqrt{2}v_Lf_L, M_{RR}=\sqrt{2}v_Rf_R,
	\end{equation}
	\begin{equation}\label{eq8}
	\rm h_D=\frac{(k_1 y+k_2 y')}{\sqrt{2} \sqrt{k^2_1+k^2_2}}.
	\end{equation}
	$M_D$, $M_{LL}$ and $M_{RR}$ being the Dirac neutrino mass matrix, left handed and right handed
	Majorana mass matrix respectively. The first and second terms in equation (\ref{eq7}) correspond to type II seesaw and type I seesaw contributions respectively. If we consider the first term on the right hand side of the above expression, one can make an estimate of neutrino mass for TeV scale LRSM. Considering $v_R \sim 6$ TeV, the type II seesaw term will be of the order of light neutrino mass $M_{\nu} \sim 0.1$ eV if 
$$ \gamma \approx \frac{5.6 \times 10^{-7}}{M_1} $$
where $M_1$ is the right handed neutrino mass. Thus, for TeV scale right handed neutrino masses, the dimensionless parameter $\gamma$ needs to be fine-tuned in order to get correct order of neutrino masses. Similar fine tuning is involved in the type I seesaw term for TeV scale $M_{RR}$. The Dirac Yukawa couplings should be fine tuned to around $10^{-6}-10^{-5}$ in order to get light neutrino mass of order $0.1$ eV. One can avoid such fine tunings if there exists structural cancellations between the two seesaw terms \cite{Luo:2008rs}.

The $6\times 6$ neutral lepton mass matrix can be diagonalised by a $6\times 6$ unitary matrix, as follows,
	\begin{equation}\label{eq14}
	\mathcal{V}^T M \mathcal{V}=\left[\begin{array}{cc}
	\widehat{M_\nu}&0\\
	0&\widehat{M}_{RR}
	\end{array}\right],
	\end{equation}
	where, $\mathcal{V}$ represents the diagonalising matrix of the full neutrino mass matrix, $M$, $\widehat{M_\nu}= {\rm diag}(m_1,m_2,m_3)$, with $m_i$ being the light neutrino masses and 
	$\widehat{M}_{RR}= {\rm diag}(M_1,M_2,M_3)$, with $M_i$ being the heavy right handed neutrino masses. $\mathcal{V}$ is thus represented as,
	\begin{equation}\label{eq15}
	\mathcal{V}=\left[\begin{array}{cc}
	U&S\\
	T&V
	\end{array}\right] \approx \left[\begin{array}{cc}
	1-\frac{1}{2}RR^{\dagger}&R\\
	-R^\dagger&1-\frac{1}{2}R^\dagger R
	\end{array}\right] \left[\begin{array}{cc}
	V_\nu&0\\
	0&V_R
	\end{array}\right],
	\end{equation}
	where, R describes the left-right mixing and given by,
	\begin{equation}\label{eq16}
	R=M_D M^{-1}_{RR}+\mathcal{O} (M^3_D{(M^{-1}_{RR})}^3).
	\end{equation}
	The matrices U, V, S and T are as follows,
	\begin{equation}\label{eq17}
	U=\left[1-\frac{1}{2}M_DM^{-1}_{RR}{(M_DM^{-1}_{RR})}^\dagger\right]V_\nu,V=\left[1-\frac{1}{2}{(M_DM^{-1}_{RR})}^\dagger M_DM^{-1}_{RR}\right]V_R
	\end{equation}
	\begin{equation}
	S=M_DM^{-1}_{RR}V_R, T=-(M_DM^{-1}_{RR})^\dagger V_\nu.
	\label{hlmixing1}
	\end{equation}
Thus, the heavy-light neutrino mixing can be parametrised in terms of the mixing matrix $T$ denoted above. This can be achieved by appropriate tuning of $M_D$ and $M_{RR}$ without crucially affecting light neutrino mass and mixing which comes not only from type I seesaw contribution involving $M_D$ and $M_{RR}$ but from type II seesaw term as well. Therefore, any fine-tuning associated with $M_D$, in order to achieve desired heavy-light mixing, can be compensated by type II seesaw term. Such heavy-light mixing has crucial significance for keV scale right handed neutrino DM, as we discuss later.

\section{Dark Matter}
\label{sec2}
The minimal LRSM discussed above does not have a stable cold dark matter candidate. One can however, minimally extend the model by including additional scalar or fermionic multiplets in the spirit of minimal dark matter scenario \cite{Cirelli:2005uq,Garcia-Cely:2015dda,Cirelli:2015bda}. Such minimal dark matter scenario in LRSM has been studied recently by the authors of \cite{Heeck:2015qra,Garcia-Cely:2015quu}. In these models, the dark matter candidate is stabilised either by a $Z_2 = (-1)^{B-L}$ subgroup of the $U(1)_{B-L}$ gauge symmetry or due to an accidental symmetry at the renormalisable level due to the absence of any renormalisable operator leading to dark matter decay \cite{Borah:2016hqn}. Some more studies on left-right dark matter also appeared in the recent works \cite{Borah:2016ees, Berlin:2016eem, Borah:2016lrl, Borah:2017leo, Borah:2017xgm}. The possibility of right handed neutrino dark matter in a different version of LRSM where the right handed lepton doublets do not contain the usual charged leptons, was also studied in the recent works \cite{Dev:2016qbd, Dev:2016xcp, Dev:2016qeb}. In our framework, the CDM candidates are stabilised due to a remnant $$\mathcal{Z}_2\simeq (-1)^{B-L}$$ 
symmetry arising after the spontaneous symmetry breaking of LRSM down to SM gauge group i.e, $SU(2)_R \times U(1)_{B-L} \to U(1)_Y$. Under this 
remnant discrete symmetry $\mathcal{Z}_2\simeq (-1)^{B-L}$, the usual leptons are odd while all bosons including scalars and gauge bosons are even. Since fermion triplets have vanishing $B-L$ charge, they are even under the remnant discrete symmetry, prohibiting them from decaying into SM leptons which are $\mathcal{Z}_2$ odd. Similarly, scalar doublets having $B-L$ charges unity are odd under this remnant discrete symmetry and hence can be stable.

The possibility of warm dark matter within minimal LRSM was also studied in the works \cite{Nemevsek:2012cd, Bezrukov:2009th}. In these works, the lightest right handed neutrino with keV scale mass was considered to be the WDM candidate. Such a WDM candidate can decay into a light neutrino and a photon at one loop level and can be cosmologically long-lived if the mixing with the light active neutrinos are appropriately tuned. Such a keV scale right handed neutrino typically gets overproduced in the early Universe, by virtue of its gauge interactions with the standard model particles. In the above mentioned works, the abundance of WDM was brought to the observed DM limits by late time entropy dilution mechanism due to the late decay of heavier right handed neutrinos \cite{Scherrer:1984fd}. In such scenarios, we need to fine tune the Yukawa couplings in order to keep the mixing of WDM with light neutrinos small as well as to allow the late decay of heavier right handed neutrinos for entropy dilution. Also, if the WDM in such models are responsible for the origin of the 3.55 keV monochromatic X-ray line, then also one requires small mixing angle, requiring some amount of fine-tuning.

To have a minimal mixed dark matter scenario in this model, we can either add a pair of scalar doublets  $\eta_{L} (1,2,1,-1), \eta_R (1,1, 2,-1)$ or a pair of fermion triplets $\Sigma_L \equiv (1,3,1,0), \Sigma_R \equiv (1,1,3,0)$ to the minimal LRSM discussed above. Higher multiplets will also work, but we stick to doublet/triplet for minimality. Here, the numbers in brackets correspond to the quantum numbers under the gauge symmetry of the model $SU(3)_c \times SU(2)_L \times SU(2)_R \times U(1)_{B-L}$. The neutral components of these multiplets can be good CDM candidates. While the discussion of left scalar dark matter is similar to the inert scalar doublet model studied extensively in the literature \cite{Ma:2006km,Barbieri:2006dq,Cirelli:2005uq, LopezHonorez:2006gr, Honorez:2010re, LopezHonorez:2010tb, Borah:2012pu, Arhrib:2013ela, Dasgupta:2014hha, Borah:2017dqx, Borah:2017dfn}, the right handed scalar dark matter $\eta^0_R$ was studied recently by the authors of \cite{Garcia-Cely:2015quu, Borah:2016hqn, Borah:2017leo}. Introducing such additional scalar doublets brings additional terms in the scalar potential (shown in appendix \ref{appen1}) given by
\begin{equation}
V_{\rm new} = V_{\eta} + V_{\Phi \eta}+ V_{\Delta \eta}.
\end{equation}
The details of different terms on the right-hand side of the above equation can be written as follows,
\begin{align}
V_{\eta} = ~ & \mu^2_{\eta} (\eta_L^{\dagger} \eta_L +\eta_R^{\dagger} \eta_R) + \rho_5 (\left[\eta_L^{\dagger} \eta_L\right]^2 +\left[\eta_R^{\dagger} \eta_R\right]^2) \nonumber \\
& +\rho_6 \left[\eta_L^{\dagger} \eta_L\right]\left[\eta_R^{\dagger} \eta_R\right],
\end{align}
\begin{align}
 V_{\Phi \eta}  = ~
& \mu_{14}\eta^\dagger_L \Phi \eta_R+f_{145} \Tr{\Phi^{\dagger} \Phi} (\eta_L^{\dagger} \eta_L+\eta_R^{\dagger} \eta_R),
 \end{align}
\begin{align}
 V_{\Delta \eta}  = ~
& (\mu_{15}\eta_L \Delta_L \eta_L+ \mu_{16}\eta_R \Delta_R \eta_R+\text{h.c.})+f_{145} (\Tr{\Delta^{\dagger}_L \Delta_L}+\Tr{\Delta^{\dagger}_R \Delta_R}) (\eta_L^{\dagger} \eta_L+\eta_R^{\dagger} \eta_R) 
 \end{align}
As such scalar interactions suggest, both left and right scalar doublets can not be dark matter at the same time as the heavier one can decay into the lighter one by virtue of their couplings to the scalar bidoublet (like the trilinear interactions of the form $\mu_{14}\eta^\dagger_L \Phi \eta_R$). Apart from such scalar interactions, the interactions of such scalars with gauge bosons can also play non trivial roles in dark matter calculations. Such interactions originate from the respective kinetic terms written in terms of appropriate covariant derivatives for LRSM gauge symmetry. The covariant derivative for the gauge group of
Left-Right model can be written as:
\begin{align}
    D^\mu_{L,R} &= \left(\partial_\mu - i g_{L,R} \frac{\vec{\tau}}{2}\vec{W}_{L,R} - ig_{B-L}\frac{({\bf B-L})}{2}B_\mu\right)
\end{align}
Now, from the above covariant derivative the kinetic part of the Lagrangian for fermion doublets $\psi$, $\psi'$ and scalar doublets $H_L, H_R$ are as follows:
\begin{align}
    \mathcal{L}_{\rm kin} \subset  (D_{\mu R} H_R)^\dagger(D^\mu_R H_R) + (D_{\mu L} H_L)^{\dagger}(D^\mu_L H_L)
\end{align}
While we consider both scalar and gauge interactions for the left scalar doublet dark matter, we consider only the gauge interactions for right scalar doublet dark matter in this work, an approach which was also adopted in earlier works \cite{Garcia-Cely:2015quu, Borah:2016hqn, Borah:2017leo}. This enables us to constrain the right handed gauge sector more from the requirement of correct relic abundance of right scalar doublet dark matter.

The left fermion triplet dark matter was studied a few years back \cite{Ma:2008cu} whereas the right fermion dark matter was studied more recently within the context of LRSM in \cite{Heeck:2015qra, Garcia-Cely:2015quu, Borah:2016ees, Bandyopadhyay:2017uwc, Arbelaez:2017ptu, Borah:2017xgm}. Such fermion triplets $\Sigma_{L,R}$ can be written in terms of the following matrix representation 
\begin{eqnarray}
&&\Sigma_{L}=\begin{pmatrix}
  \Sigma^0_{L}  & \sqrt{2} \Sigma^+_{L}  \\
  \sqrt{2} \Sigma^-_{L} & -\Sigma^0_{L}
 \end{pmatrix} \equiv [3,1,0,1] \, , \nonumber \\
&&\Sigma_{R}=\begin{pmatrix}
  \Sigma^0_{R}  & \sqrt{2} \Sigma^+_{R}  \\
  \sqrt{2} \Sigma^-_{R} & -\Sigma^0_{R}
 \end{pmatrix} \equiv [1,3,0,1] \, 
\end{eqnarray}
where the neutral component of the each fermion triplet can be a stable dark matter candidate. As discussed in these works, the phenomenology of fermion triplet CDM is much richer due to the fact that both the left and right handed fermion can be stable giving rise to a multi-component CDM scenario. Also, the fermion triplet dark matter related calculations are simpler as they are mostly governed by their respective kinetic terms. This is due to the absence of any interactions with the scalar fields at the renormalisable level. The kinetic terms of the fermion triplets lead to the following interactions with the gauge bosons
\begin{align}
\begin{split}
\L_\Sigma &\supset  \big[ g_L    \overline{\Sigma}^+_L \slashed{W}_{L}^3 \Sigma^+_L  + \sqrt{2} g_L  \overline{\Sigma}_L^{+} \slashed{W}^+_L \Sigma^0_L + \hc \big] \\
&+\big[ g_R    \overline{\Sigma}^+_R \slashed{W}_{R}^3 \Sigma^+_R  + \sqrt{2} g_R  \overline{\Sigma}_R^{+} \slashed{W}^+_R \Sigma^0_R+ \hc \big] \, .
\end{split}
\end{align}
Since the high $SU(2)$ dimensions of triplet fermions do not allow them to couple to fermions and scalars, the only interactions affecting relic abundance are the gauge interactions from the above kinetic terms.

While the lightest right handed neutrino having mass in keV regime can naturally be a WDM candidate \cite{Nemevsek:2012cd, Bezrukov:2009th}, one can also have a more exotic version of the LRSM where the warm dark matter component can be a fundamental scalar. Although there involves an issue of fine-tuning in generating keV scale or smaller mass of a scalar, there are some advantages of this scenario. Firstly, the lower bound on dark matter mass is not applicable like it is there in case of fermion DM from the galactic phase space criteria. If a fermion DM is to constitute the entire DM in a galaxy, then below a certain mass, the phase space density of DM particles that would be required by the observed amount of DM in dwarf galaxies, would violate the Pauli exclusion principle. This lower bound on fermion DM mass (around 0.4 keV) was calculated long back by Tremaine-Gunn \cite{Tremaine:1979we}. For scalar DM, this bound is relaxed as the Pauli exclusion principle does not apply there. Since the scalar doublets $\eta_{L, R}$ can not give rise to two stable or long-lived DM candidates as mentioned earlier, we therefore introduce a pair of scalar triplets $\Omega_L \equiv (1,3,1,0), \Omega_R \equiv (1,1,3,0)$. In such a setup, the lighter of $\eta^0_{L, R}$ can be CDM while $\Omega^0_R$ can be WDM. As mentioned earlier, we confine ourselves to right handed sector for WDM in order to avoid precision constraints due to keV scale particles having electroweak gauge interactions.

\subsection{Relic Abundance Calculation of CDM}
Several astrophysical and cosmological evidences suggest the presence of dark matter (DM) in our Universe. The latest data collected by the Planck experiment suggests around $26\%$ of the present Universe's energy density being made up of dark matter \cite{Ade:2015xua} as mentioned earlier in \eqref{dm_relic}. According to the list of criteria, a dark matter candidate must fulfil \cite{Taoso:2007qk}, none of the SM particles can qualify for it. In this section, we outline the standard procedures to calculate the abundance of both keV and TeV-ish DM candidates.
The relic abundance of a dark matter particle $\rm DM$, which was in thermal equilibrium at some earlier epoch can be calculated by solving the Boltzmann equation
\begin{equation}
\frac{dn_{\rm DM}}{dt}+3Hn_{\rm DM} = -\langle \sigma v \rangle (n^2_{\rm DM} -(n^{\rm eq}_{\rm DM})^2)
\end{equation}
where $n_{\rm DM}$ is the number density of the dark matter particle $\rm DM$ and $n^{\rm eq}_{\rm DM}$ is the number density when $\rm DM$ was in thermal equilibrium. $H$ is the Hubble expansion rate of the Universe and $ \langle \sigma v \rangle $ is the thermally averaged annihilation cross section of the dark matter particle $\rm DM$. In terms of partial wave expansion $ \langle \sigma v \rangle = a +b v^2$. Numerical solution of the Boltzmann equation above gives \cite{Kolb:1990vq,Scherrer:1985zt}
\begin{equation}
\Omega_{\rm DM} h^2 \approx \frac{1.04 \times 10^9 x_F}{M_{\text{Pl}} \sqrt{g_*} (a+3b/x_F)}
\end{equation}
where $x_F = M_{\rm DM}/T_F$, $T_F$ is the freeze-out temperature, $M_{\rm DM}$ is the mass of dark matter, $g_*$ is the number of relativistic degrees of freedom at the time of freeze-out and and $M_{\text{Pl}} \approx 2.4\times 10^{18}$ GeV is the Planck mass. Dark matter particles with electroweak scale mass and couplings freeze out at temperatures approximately in the range $x_F \approx 20-30$. More generally, $x_F$ can be calculated from the relation 
\begin{equation}
x_F = \ln \frac{0.038gM_{\text{Pl}}M_{\rm DM}<\sigma v>}{g_*^{1/2}x_F^{1/2}}
\label{xf}
\end{equation}
which can be derived from the equality condition of DM interaction rate $\Gamma = n_{\rm DM} \langle \sigma v \rangle$ with the rate of expansion of the Universe $H \approx g^{1/2}_*\frac{T^2}{M_{Pl}}$. There also exists a simpler analytical formula for the approximate DM relic abundance \cite{Jungman:1995df}
\begin{equation}
\Omega_{\rm DM} h^2 \approx \frac{3 \times 10^{-27} cm^3 s^{-1}}{\langle \sigma v \rangle}
\label{eq:relic}
\end{equation}
The thermal averaged annihilation cross section $\langle \sigma v \rangle$ is given by \cite{Gondolo:1990dk}
\begin{equation}
\langle \sigma v \rangle = \frac{1}{8M^4_{\rm DM}T K^2_2(M_{\rm DM}/T)} \int^{\infty}_{4M^2_{\rm DM}}\sigma (s-4M^2_{\rm DM})\surd{s}K_1(\surd{s}/T) ds
\end{equation}
where $K_i$'s are modified Bessel functions of order $i$ and $T$ is the temperature.

If there exists some additional particles having masses close to that of DM, then they can be thermally accessible during the epoch of DM freeze out. The can give rise to additional channels through which DM can coannihilate with such additional particles and produce SM particles in the final states. This type of coannihilation effects on dark matter relic abundance were studied by several authors in \cite{Griest:1990kh, Edsjo:1997bg, Bell:2013wua}. Here we summarise the analysis of \cite{Griest:1990kh} for the calculation of the effective annihilation cross section in such a case. The effective cross section can given as 
\begin{align}
\sigma_{\rm eff} &= \sum_{i,j}^{N}\langle \sigma_{ij} v\rangle r_ir_j \nonumber \\
&= \sum_{i,j}^{N}\langle \sigma_{ij}v\rangle \frac{g_ig_j}{g^2_{\rm eff}}(1+\Delta_i)^{3/2}(1+\Delta_j)^{3/2}e^{\big(-x_F(\Delta_i + \Delta_j)\big)} \nonumber \\
\end{align}
where $x_F = \frac{M_{\rm DM}}{T_F}$ and $\Delta_i = \frac{m_i-M_{\text{DM}}}{M_{\text{DM}}}$ and 
\begin{align}
g_{\rm eff} &= \sum_{i=1}^{N}g_i(1+\Delta_i)^{3/2}e^{-x_F\Delta_i}
\end{align}
with $g_{i, j}$ being the internal degrees of freedom for species $i, j$ respectively and N is the number of coannihilating particles during the epoch of DM freeze-out. The masses of the heavier components of the inert Higgs doublet are denoted by $m_{i}$. The thermally averaged cross section between two coannihilating particles $i, j$ with masses $m_i, m_j$ can be written as
\begin{align}
\langle \sigma_{ij} v \rangle &= \frac{x_F}{8m^2_im^2_jM_{\text{DM}}K_2((m_i/M_{\text{DM}})x_F)K_2((m_j/M_{\text{DM}})x_F)} \times \nonumber \\
& \int^{\infty}_{(m_i+m_j)^2}ds \sigma_{ij}(s-2(m_i^2+m_j^2)) \sqrt{s}K_1(\sqrt{s}x_F/M_{\text{DM}}) \nonumber \\
\label{eq:thcs}
\end{align}
We use micrOMEGAs \cite{Belanger:2013oya} to compute the relic abundance of CDM in our work.

\subsection{Relic Abundance Calculation of Thermal WDM}
The lightest right handed neutrino can be long lived if it has a mass below the mass of an electron, since it can decay only at loop level into lighter particles like standard model neutrinos and photon. Since the right handed neutrino $N_1$ has gauge interactions in LRSM, they can be in thermal equilibrium in the early Universe freezing out subsequently around 
\begin{equation}
T_{fN_1} \approx g^{1/6}_{*f} \left (\frac{M_{W_R}}{M_{W_L}} \right)^{4/3} T_{f\nu}
\label{nuRfreeze}
\end{equation}
with $g_{*f}$ being the relativistic degrees of freedom at $T=T_{fN_1}$. It is defined as
$$ g_{*}= \sum_{i \in \text{boson}} \left ( \frac{T_i}{T} \right)^4 g_i + \frac{7}{8} \sum_{i \in \text{fermion}} \left( \frac{T_i}{T} \right)^4 g_i. $$
If all relativistic particles are in equilibrium with each other, it can simply be written as
$$ g_{*}= \sum_{i \in \text{boson}} g_i + \frac{7}{8} \sum_{i \in \text{fermion}} g_i. $$
In the above equation \eqref{nuRfreeze}, $T_{f\nu} \sim 1-2$ MeV is the freeze-out temperature of light neutrinos. Thus, for TeV scale $W_R$, the right handed neutrino can remain in equilibrium until late epochs corresponding to a temperature of a few hundred MeV's. At such high temperatures, a keV right handed neutrino can behave like a relativistic species whose number and entropy densities can be given as
\begin{equation}
n=g_{n*} \frac{\zeta(3)}{\pi^2} g_i T^3, \;\;\;\; s=\frac{2\pi^4}{45} g_{\text{eff}} T^3
\end{equation}
where $g_{n*} = 1, \frac{3}{4}$ for boson, fermion respectively, and $g_{\text{eff}}$ is given by
$$ g_{\text{eff}}= \sum_{i \in \text{boson}} \left ( \frac{T_i}{T} \right)^3 g_i + \frac{7}{8} \sum_{i \in \text{fermion}} \left( \frac{T_i}{T} \right)^3 g_i. $$
Before QCD phase transition temperature ($\sim$ a few hundred MeV), since all relativistic species are in equilibrium with each other $(T_i = T, \forall i)$ we can write the effective relativistic degrees of freedom for entropy density as 
$$ g_{\text{eff}}= \sum_{i \in \text{boson}} g_i + \frac{7}{8} \sum_{i \in \text{fermion}} g_i. $$
The $N_1$ number density to entropy density after freeze-out is given by 
\begin{equation}
\frac{n_{N_1}}{s}\rvert_f = \frac{1}{g_{*f}} \frac{135 \zeta(3)}{4\pi^4}
\end{equation}
The present abundance of $N_1$ in comparison to the total DM abundance is
\begin{equation}
\frac{\Omega_{N_1}}{\Omega_{\text{DM}}} = \frac{n_{N_1}}{s} \rvert_f \frac{M_{N_1} s_0}{\Omega_{\text{DM}} \rho_c} \approx \frac{1}{g_{*f}} 7.59 \times 10^3 \left (\frac{M_{N_1}}{7 \; \text{keV}} \right)
\end{equation}
with $s_0, \rho_c= \frac{3H^2_0}{8\pi G}$ being the entropy density and critical density of the present Universe. Thus, even if the freeze-out occurs above the electroweak symmetry breaking so that $g_{*f} \approx 107$, the abundance of $N_1$ will be much more than the observed DM, overclosing the Universe. For decoupling temperature of a few hundred GeVs for which $g_{*f} \approx 60$, we can normalise the abundance of $N_1$ as
\begin{equation}
\frac{\Omega_{N_1}}{\Omega_{\text{DM}}} \approx  1.265 \times 10^2 \left( \frac{60}{g_{*f}} \right) \left (\frac{M_{N_1}}{7 \; \text{keV}} \right)
\end{equation}

This requires entropy dilution after freeze-out to bring down the abundance of $\Omega_{N_1} \leq \Omega_{\text{DM}}$. Late decay of heavier right handed neutrinos like $N_2$ can release such entropy. Such a decay should however occur before the big bang nucleosynthesis (BBN) temperature $T_{\text{BBN}} \sim \mathcal{O}$(MeV) in order to be consistent with successful BBN predictions. Such late decay of long lived particles can release extra entropy and dilute the abundance of keV dark matter to bring it into the observed limit \cite{Scherrer:1984fd}. The dilution factor due to the decay of such a heavy long lived particle $N_2$ is given by \cite{Scherrer:1984fd}
\begin{equation}
d = \frac{s_{\text{before}}}{s_{\text{after}}} \approx 0.58 [g_*(T_r)]^{-1/4} \frac{\sqrt{\Gamma_{N_2} M_{\text{Pl}}}}{M_{N_2} Y_{N_2}}\;,
\end{equation}
where $\Gamma_{N_2}$ is the decay width of the heavy particle with mass $M_{N_2}$ and
$$Y_{N_2}=\frac{n}{s} = \frac{135}{4\pi^4} \frac{\zeta(3)}{g_*(T_{fN_2})}$$
is the initial abundance of the particle $N_2$ before it started to decay. Also, $g_*(T_r)$ is the relativistic degrees of freedom at a temperature $T_r$ just after the decay of $N_2$. This temperature to which the Universe cools down to following the release of entropy due to the decay of $N_2$ can be approximated as 
\begin{equation}
T_r \approx 0.78 [g_*(T_r)]^{-1/4} \sqrt{\Gamma_{N_2} M_{\text{Pl}}}\;.
\end{equation}
Also, $g_*(T_{fN_2})$ is the relativistic degrees of freedom at the epoch of $N_2$ freeze-out. 
For maximum dilution or minimum value of $d$, it is desirable to have $g_*(T_r)$ minimum $(\approx 10.75)$, equal to the value of $g_*$ just before BBN.

Similarly, the keV scalar DM can also remain in thermal equilibrium by virtue of gauge interactions. If the neutral component of $\Omega_R$ is the keV scalar, then it has interactions with $W_R$ bosons. Since, a neutral scalar can not have three point interactions with $W_R, Z_R$ bosons, the only interactions it can have is the four point ones of the type $W^+_R W^-_R \Omega^0_R \Omega^0_R$. The interaction that can keep $\Omega^0_R$ in equilibrium until late epochs is $\gamma \gamma \rightarrow \Omega^0_R \Omega^0_R$ through a $W_R$ boson loop. The cross section can be estimated as
\begin{equation}
\sigma (\gamma \gamma \rightarrow \Omega^0_R \Omega^0_R) = \frac{E^2_{\Omega^0_R} F^2_W}{64 \pi} \left( \frac{e^2 g^2}{32 \pi^2 M^2_{W_R}} \right)^2
\end{equation}
Here $E_{\Omega^0_R} = \rho_{\Omega^0_R}/n_{\Omega^0_R} = 2.7 T$ and $F_W = 7$ is a loop function. To find the decoupling temperature, we equate the interaction rate $\Gamma$ with the Hubble expansion rate $H$ as follows.
\begin{align}
& \Gamma = n_{\gamma} \sigma v = H (T_{f\Omega}) = 1.66 \sqrt{g_{*f}}\frac{T^2_{f\Omega}}{M_{\text{Pl}}} \\
& \implies \frac{2 \zeta (3)}{\pi^2} T^3_{f\Omega}  \frac{E^2_{\Omega^0_R} F^2_W}{64 \pi} \left( \frac{e^2 g^2}{32 \pi^2 M^2_{W_R}} \right)^2=1.66 \sqrt{g_{*f}}\frac{T^2_{f\Omega}}{M_{\text{Pl}}}\nonumber \\
& \implies T_{f\Omega} = 3.58 \times 10^{-4} g^{1/6}_{*f} \left(\frac{M_{W_R}}{\text{GeV}} \right)^{4/3}
\end{align}
Thus, even if we take the lowest possible value of $M_{W_R} \sim 3$ TeV corresponding to $g_{*f}$ value of at least 107 (same as that of SM particles at high temperatures), the keV scalar DM freezes out at around $T_{f \Omega} \approx 33$ GeV. Since the scalar WDM decouples while being relativistic, it is straightforward to calculate the present abundance.

The abundance of $\Omega^0_R$ can be written in terms of the ratio of number density to entropy density as
\begin{equation}
Y_{\Omega^0_R} = \frac{n_{\Omega^0_R}}{s}
\end{equation}
Using the expressions for number and entropy densities for relativistic species, we can write it as
\begin{equation}
Y_{\Omega^0_R} = \frac{45 \zeta(3)}{2\pi^4} \frac{g_{\Omega^0_R}}{g_{\text{eff}}}
\end{equation}
Since $Y_{\Omega^0_R}$ is conserved as the Universe evolves, the present abundance can be written as 
\begin{equation}
\Omega_{\Omega^0_R} = Y_{\Omega^0_R} m_{\Omega^0_R} \frac{s_0}{\rho_c}
\end{equation}
where $s_0 \approx 2.89 \times 10^3 \; \text{cm}^{-3}$ is the entropy density and $\rho_c \approx 1.05 \times 10^{-5} h^2 \; \text{GeV} \text{cm}^{-3}$ is the critical density of the Universe at present. Using $g_{\Omega^0_R} = 1, h=0.68$ we can find
\begin{equation}
\Omega_{\Omega^0_R} = 1.645 \times 10^8 \frac{1}{g_{\text{eff}}} \left( \frac{m_{\Omega^0_R}}{\text{GeV}} \right).
\end{equation}
Using appropriate normalisations, we can rewrite it as 
\begin{equation}
\Omega_{\Omega^0_R} = 11.54 \left (\frac{100}{g_{\text{eff}}} \right) \left (\frac{m_{\Omega^0_R}}{7 \;\text{keV}} \right).
\end{equation}
Here $g_{\text{eff}} \approx 100$ is the appropriate relativistic degrees of freedom at freeze-out temperature $T_{f \Omega} \approx 33$ GeV corresponding to $M_{W_R} \sim 3$ TeV. Also, the mass of the scalar is normalised to 7 keV, which has interesting implications for the origin of 3.55 keV X-ray line as we discuss below. Thus, for this generic normalisations, the thermal abundance of keV scalar DM comes out to be around 43 times the required DM abundance. This can be brought within Planck limits by appropriate entropy dilution at late epochs. The lightest right handed neutrino $N_1$ decay can do the needful here, similar to the entropy dilution due to $N_2$ decay in the case of $N_1$ as keV DM, discussed above.

We are now going to use the standard recipe described above for calculating relic abundance of thermally produced CDM and WDM in LRSM. While earlier studied covered CDM and WDM separately, here we are going to have a mixed DM scenario comprising of both the components. Since the scale of CDM freeze-out is above the scale at which WDM decouples for typical choices of DM masses and TeV scale LRSM, we calculate their abundance separately. To be more specific, we plan to show
\begin{enumerate}
\item the change in CDM parameter space, compared to previous analysis, due to under-abundant criteria in our mixed DM scenario.
\item the change in WDM parameter space, compared to previous analysis, due to under-abundant criteria in our mixed DM scenario.
\item the change in CDM parameter space, compared to previous analysis,  due to late time entropy dilution required to bring the thermally overproduced WDM within limits.
\item the agreement with direct, indirect search as well as collider constraints and possibility of generating 3.55 keV lime from WDM decay at radiative level.
\end{enumerate}

\section{Results and Discussion}
\label{sec6}
\subsection{Relic Abundance of CDM}
We first calculate the relic abundance of different CDM candidates discussed in the work. Since we are considering a mixture of CDM and WDM, we find out the parameter space that gives rise to under-abundant CDM and compare it with the parameter space that gives $100\%$ CDM. Here we note that, the thermal abundance of CDM will also be affected by the late entropy dilution we consider to bring the overproduced WDM within Planck limits. This is because, typically CDM freezes out at temperatures corresponding to $x_F = \frac{M_{\text{CDM}}}{T_F} \approx 20-30$ whereas the diluter decays around or before BBN temperature corresponding to $1\; \rm MeV \lessapprox T < 100$ MeV. Since the entropy dilution factor for WDM is of the order of hundred, we generate those part of the CDM parameter space that overproduce it thermally, so that even after entropy dilution, some substantial amount of CDM remains in the Universe. We first find the CDM parameter space assuming the dark sector to be composed purely of CDM. Then for illustrative purposes, we constrain the parameter space for CDM corresponding to $75\%, 50\%, 25\%$ of CDM respectively in the Universe, provided the remaining fraction can arise from WDM.

\begin{figure}[!h]
\centering
\begin{tabular}{cc}
\epsfig{file=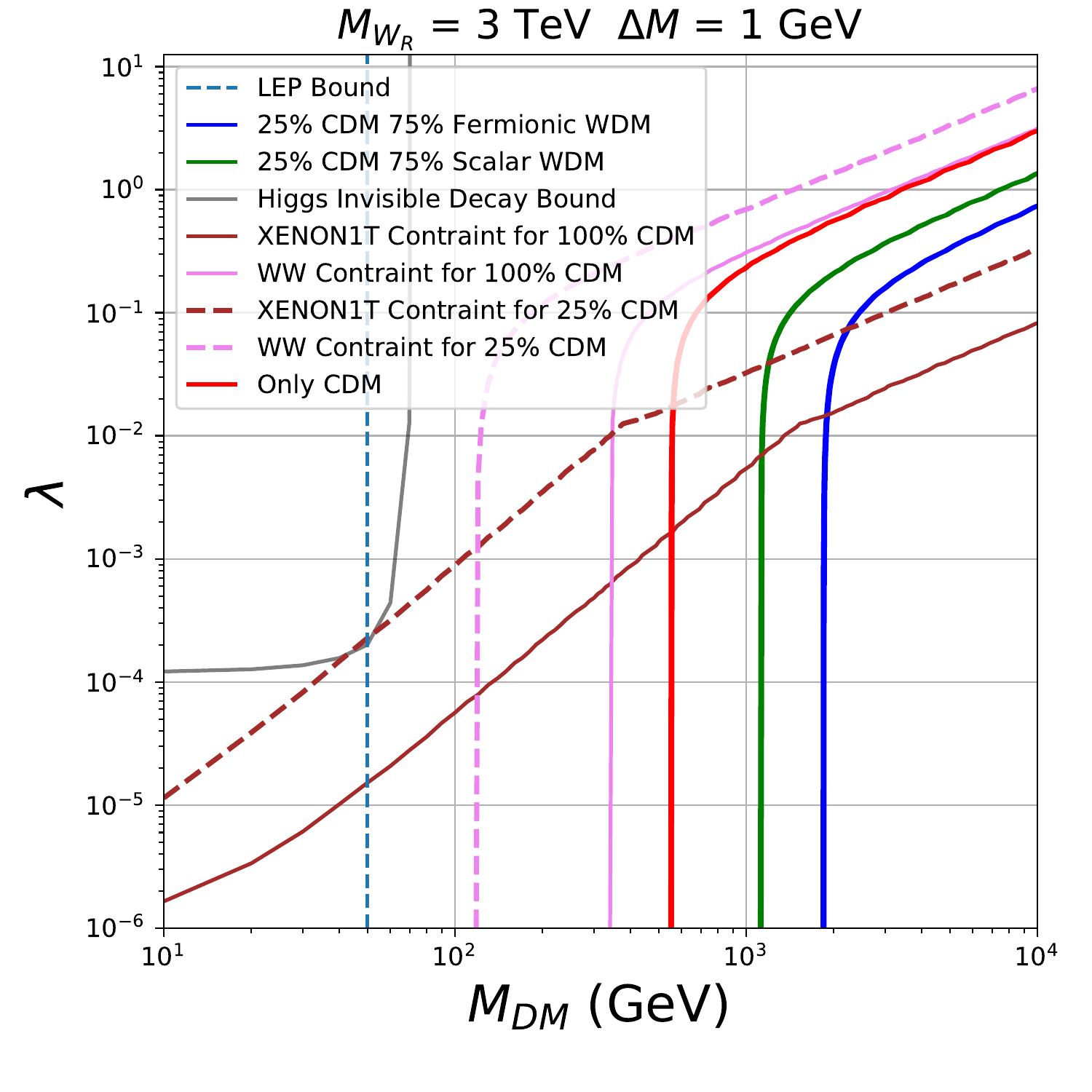,width=0.50\textwidth,clip=}
\epsfig{file=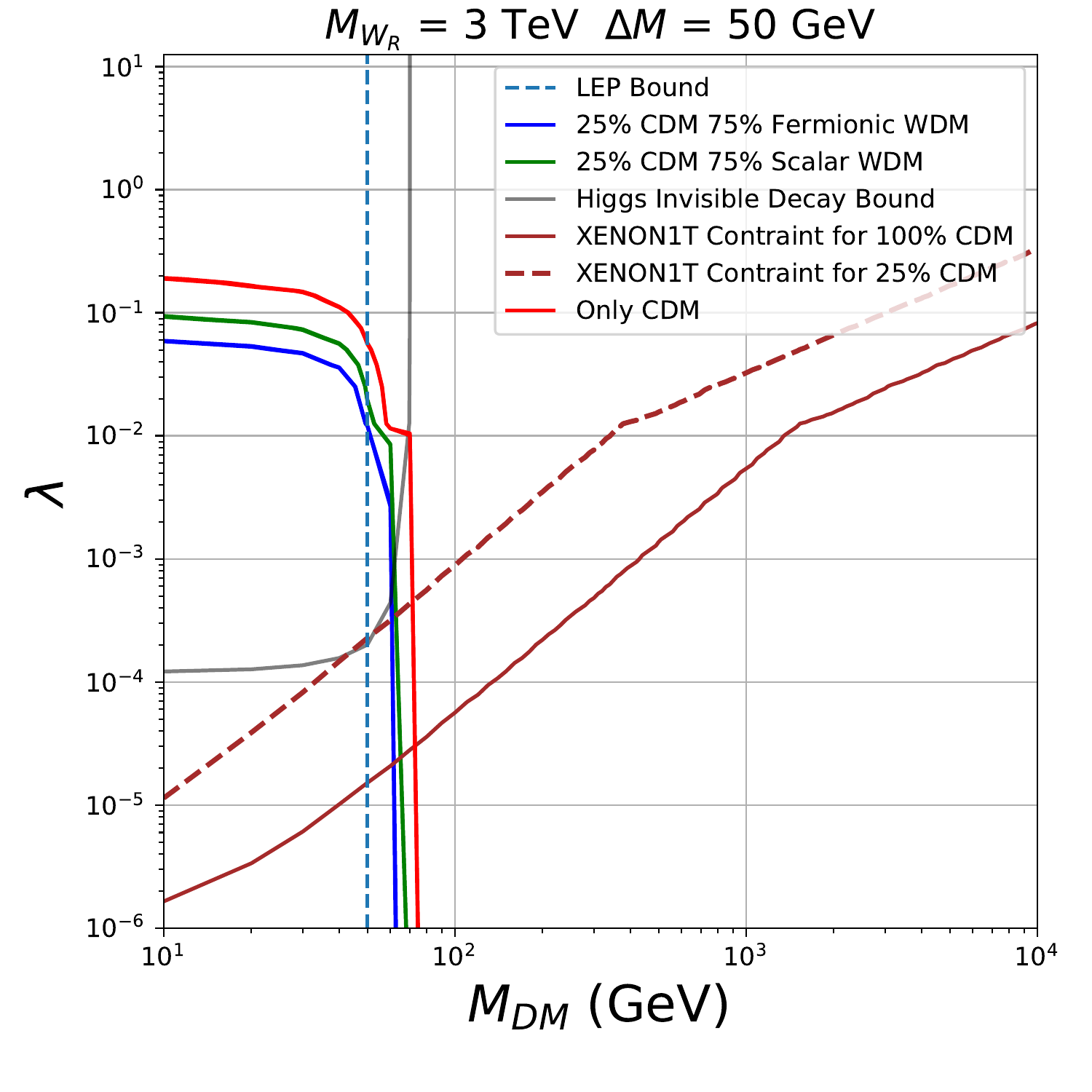,width=0.50\textwidth,clip=} \\
\epsfig{file=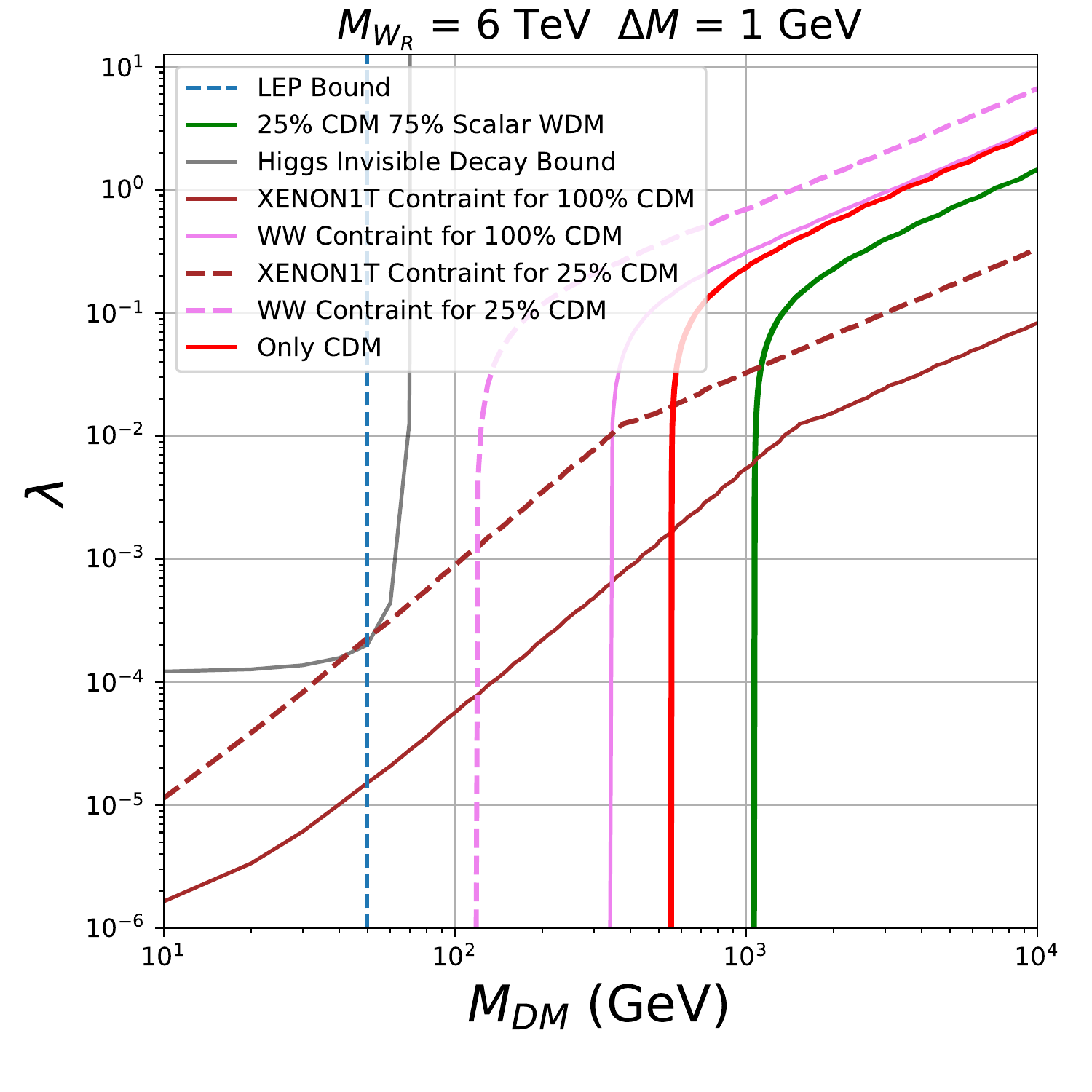,width=0.50\textwidth,clip=}
\epsfig{file=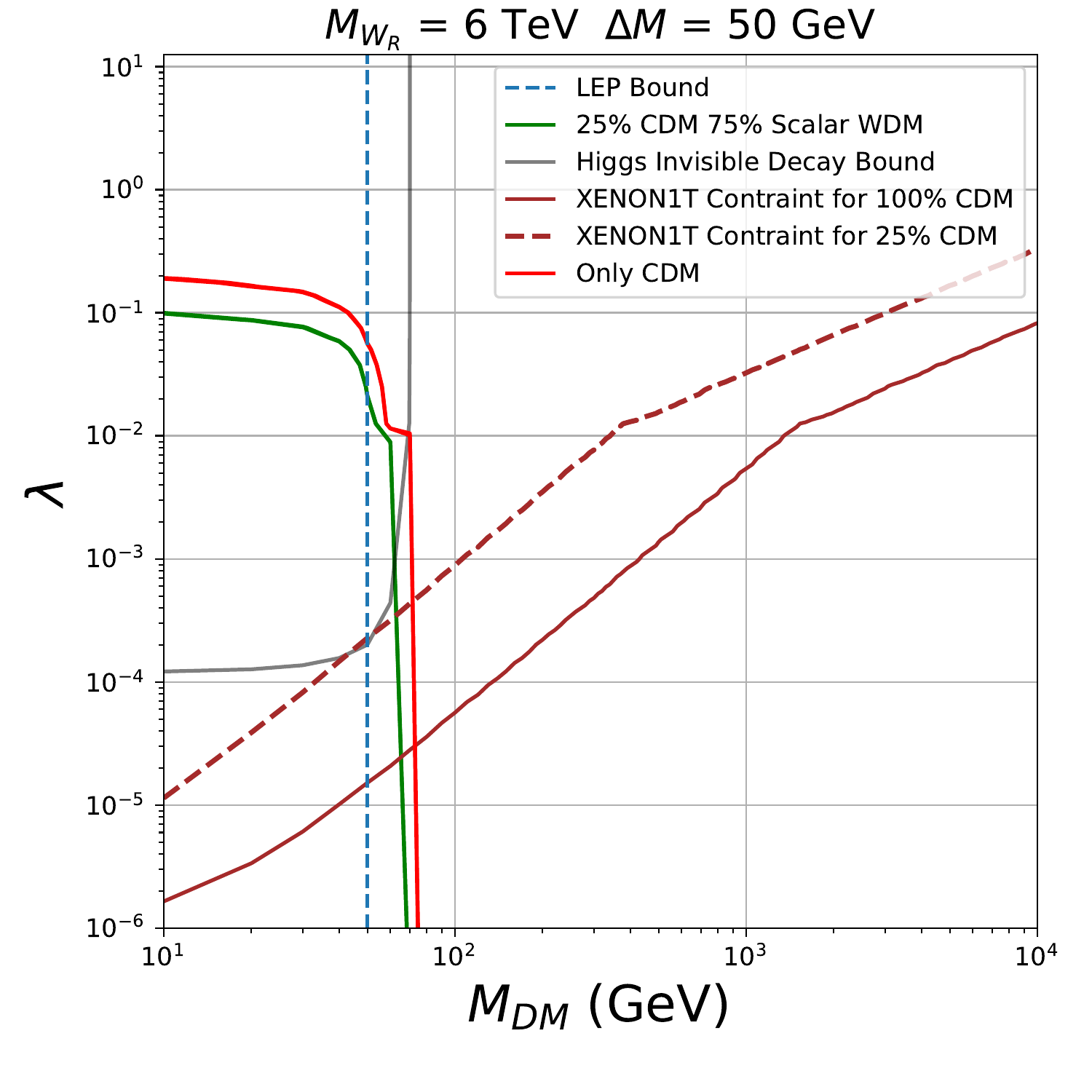,width=0.50\textwidth,clip=}
\end{tabular}
\caption{Allowed parameter space from relic abundance point of view for left scalar doublet dark matter assuming $100\%$ CDM (red solid line) and $25\%$ CDM (green solid line for scalar WDM, blue solid line for fermion WDM). The y-axis shows CDM-Higgs coupling and x-axis shows CDM mass. Choice of benchmark parameters and different relevant bounds are indicated by the legends.}
\label{fig2}
\end{figure}

\begin{figure}[!h]
\centering
\begin{tabular}{cc}
\epsfig{file=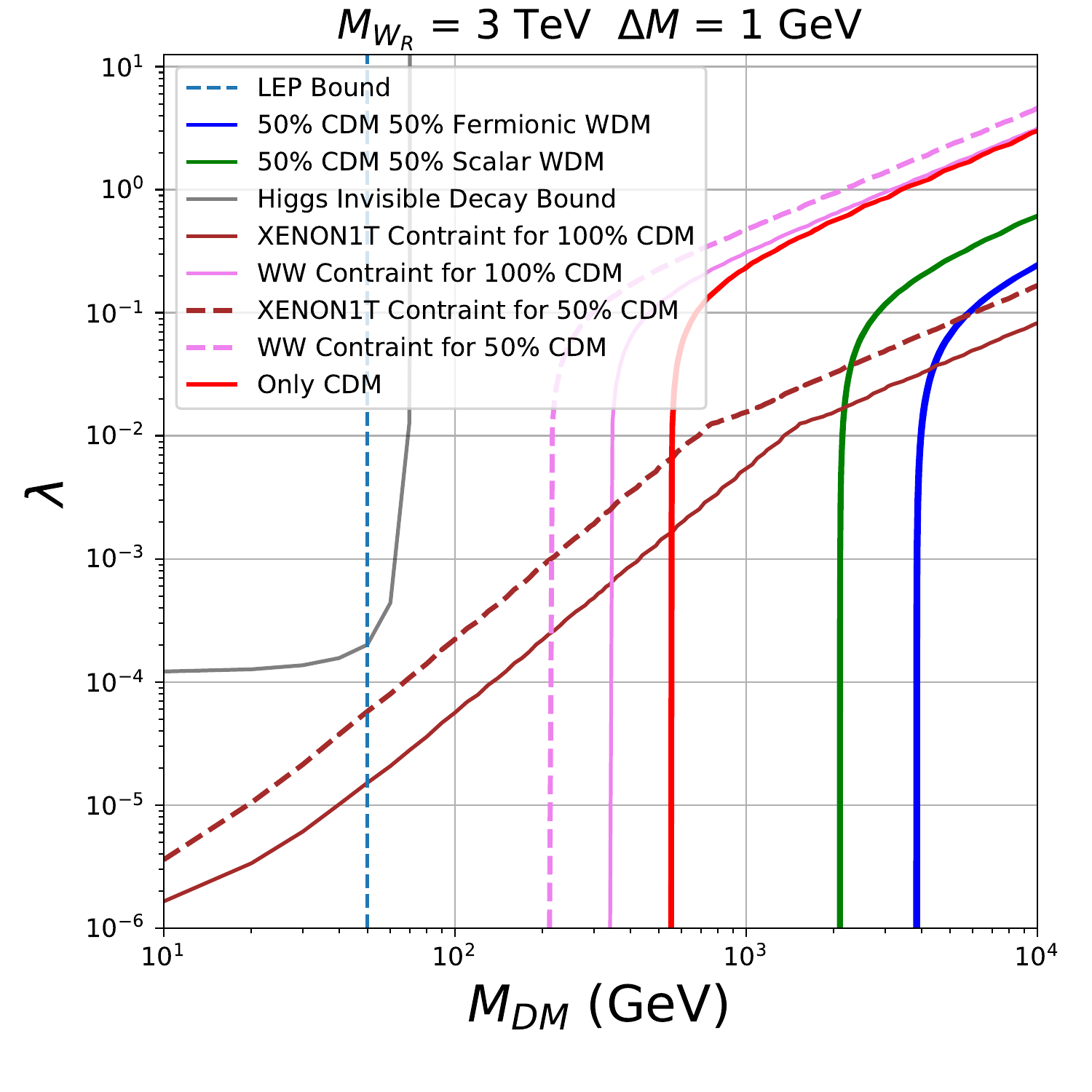,width=0.50\textwidth,clip=}
\epsfig{file=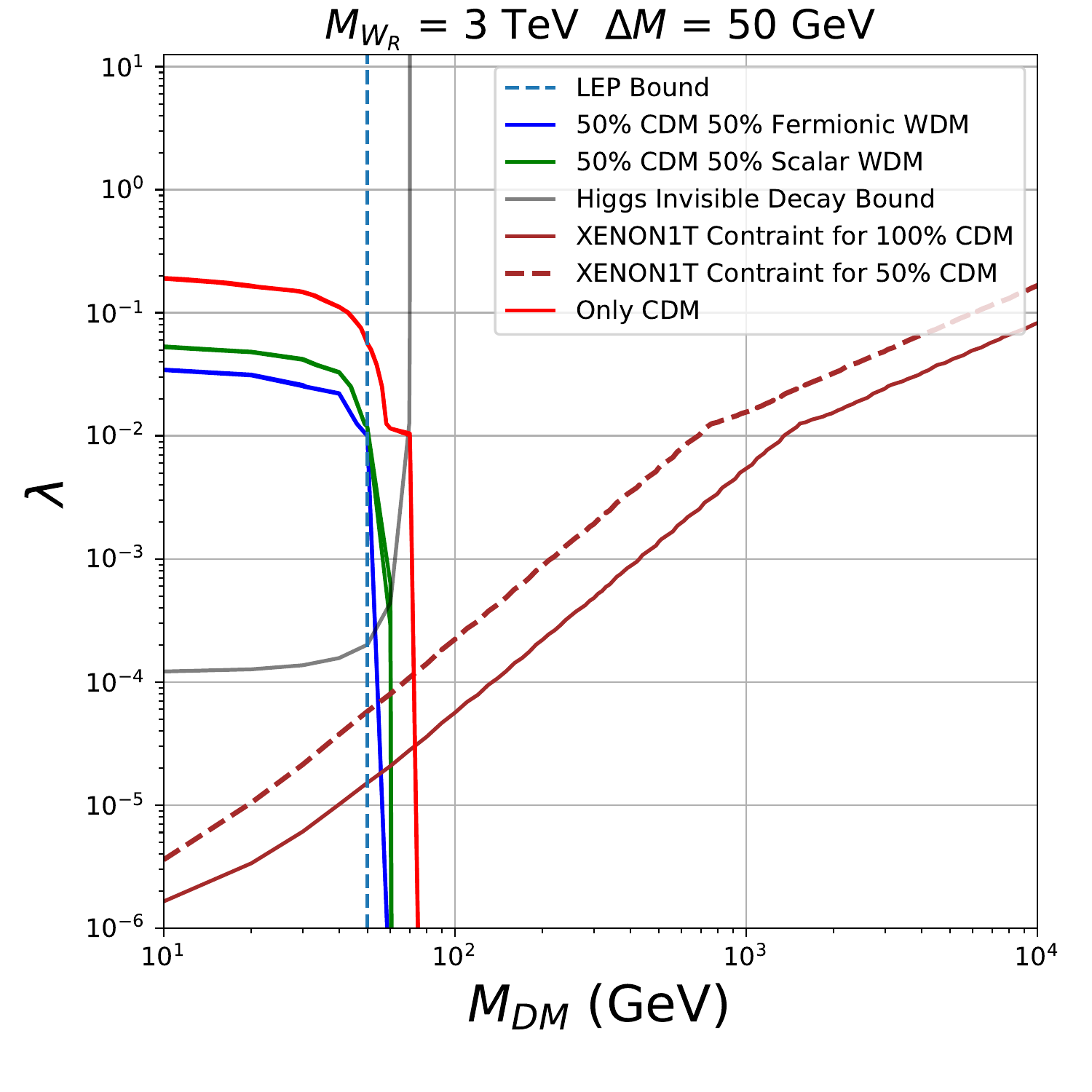,width=0.50\textwidth,clip=} \\
\epsfig{file=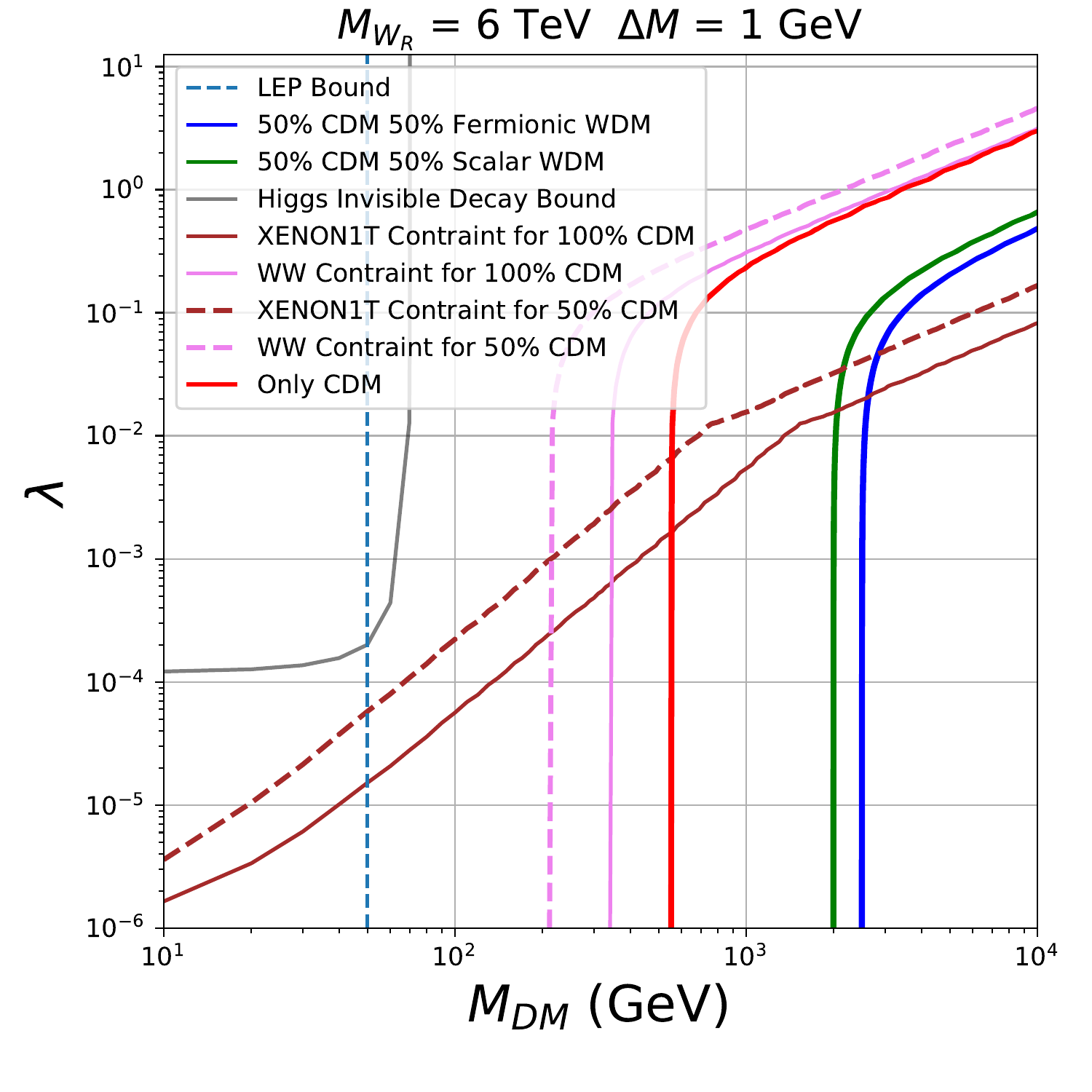,width=0.50\textwidth,clip=}
\epsfig{file=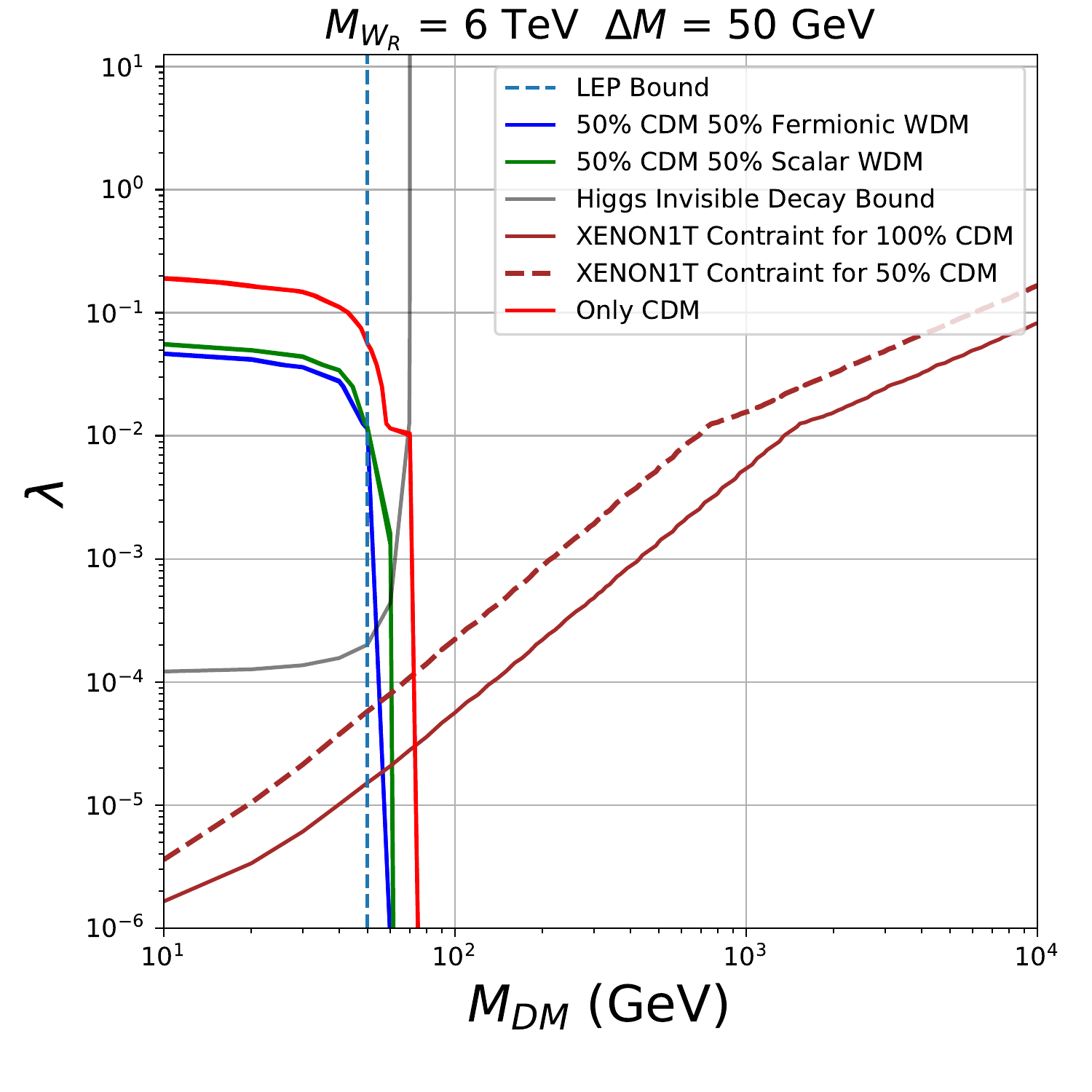,width=0.50\textwidth,clip=}
\end{tabular}
\caption{Allowed parameter space from relic abundance point of view for left scalar doublet dark matter assuming $100\%$ CDM (red solid line) and $50\%$ CDM (green solid line for scalar WDM, blue solid line for fermion WDM). The y-axis shows CDM-Higgs coupling and x-axis shows CDM mass. Choice of benchmark parameters and different relevant bounds are indicated by the legends.}
\label{fig2a}
\end{figure}

\begin{figure}[!h]
\centering
\begin{tabular}{cc}
\epsfig{file=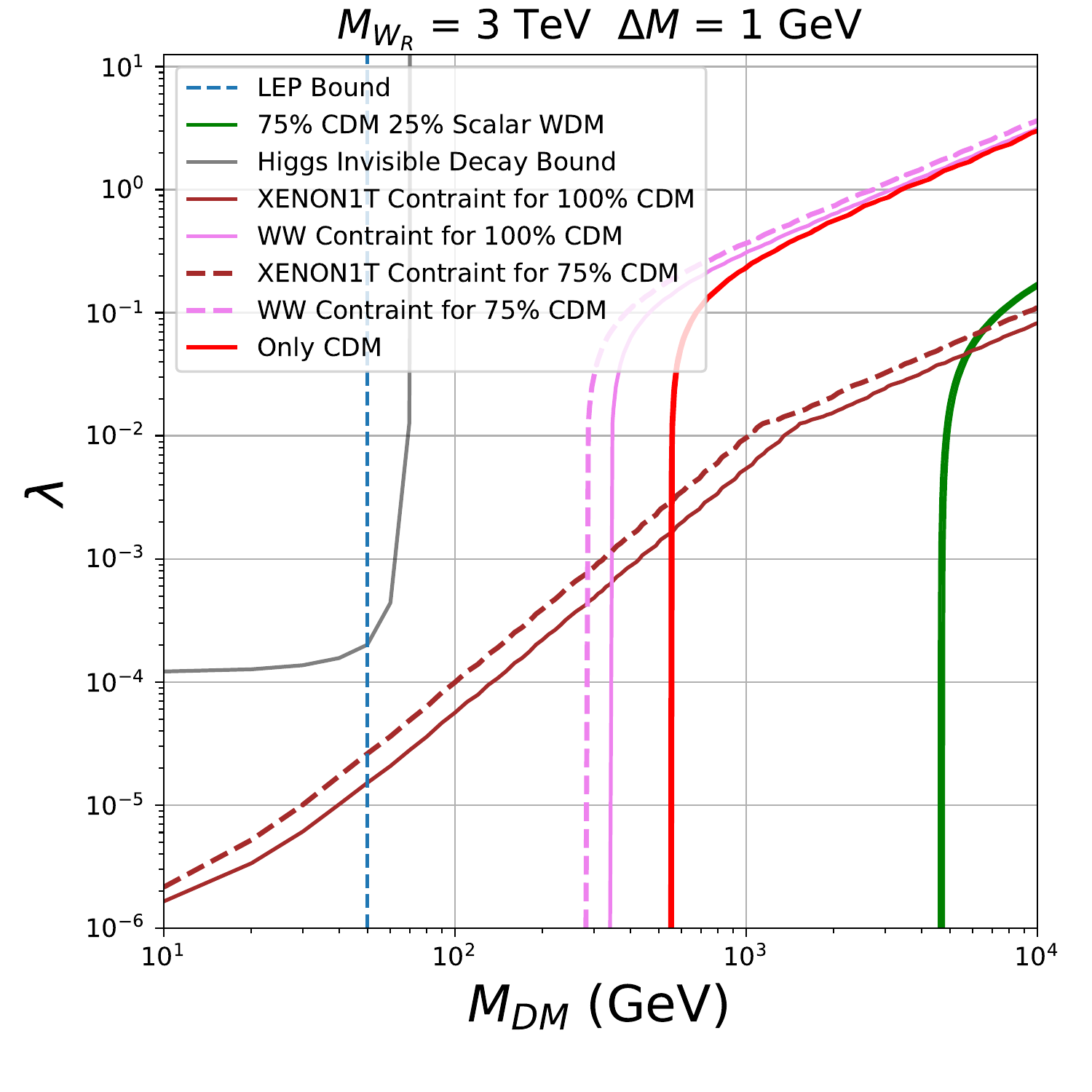,width=0.50\textwidth,clip=}
\epsfig{file=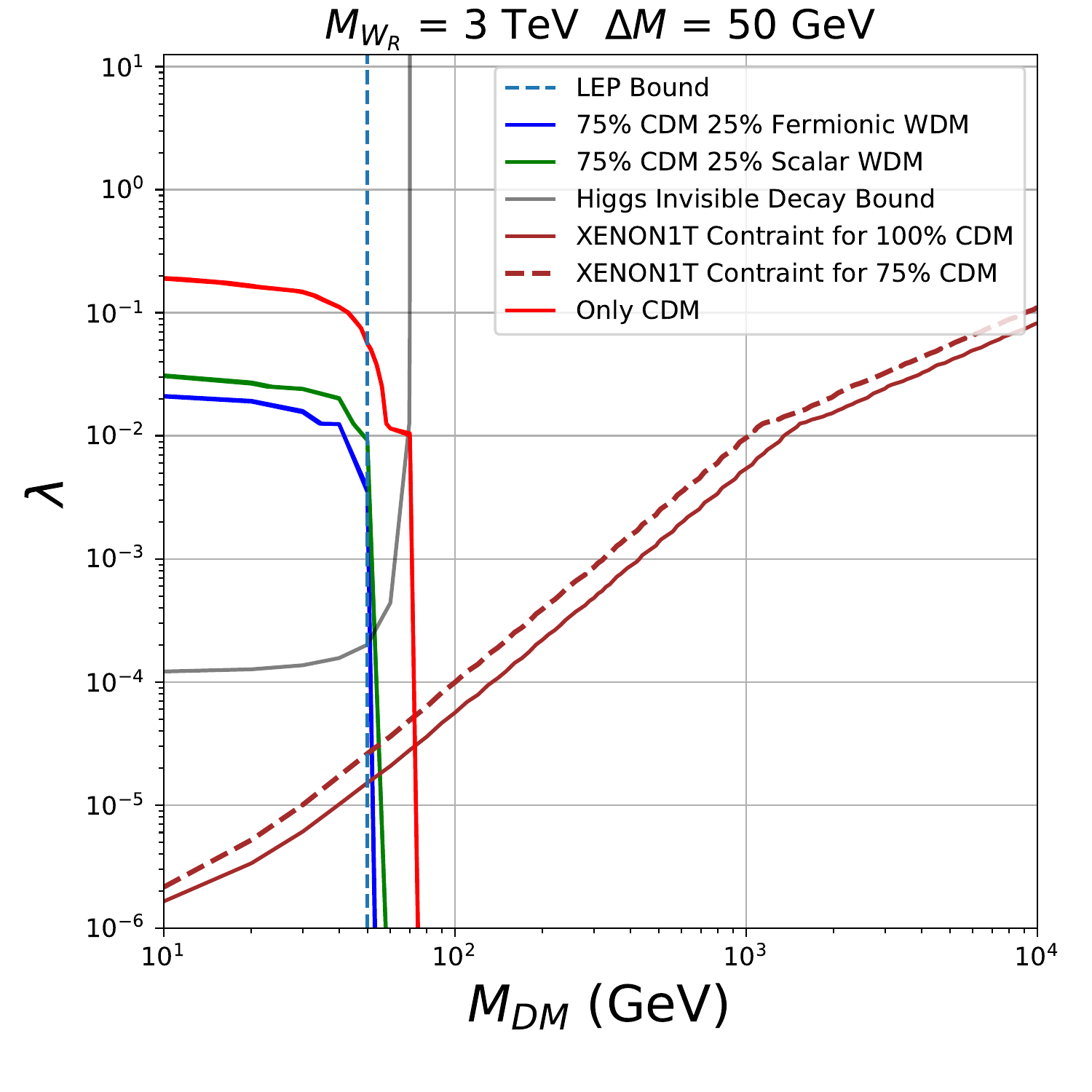,width=0.50\textwidth,clip=} \\
\epsfig{file=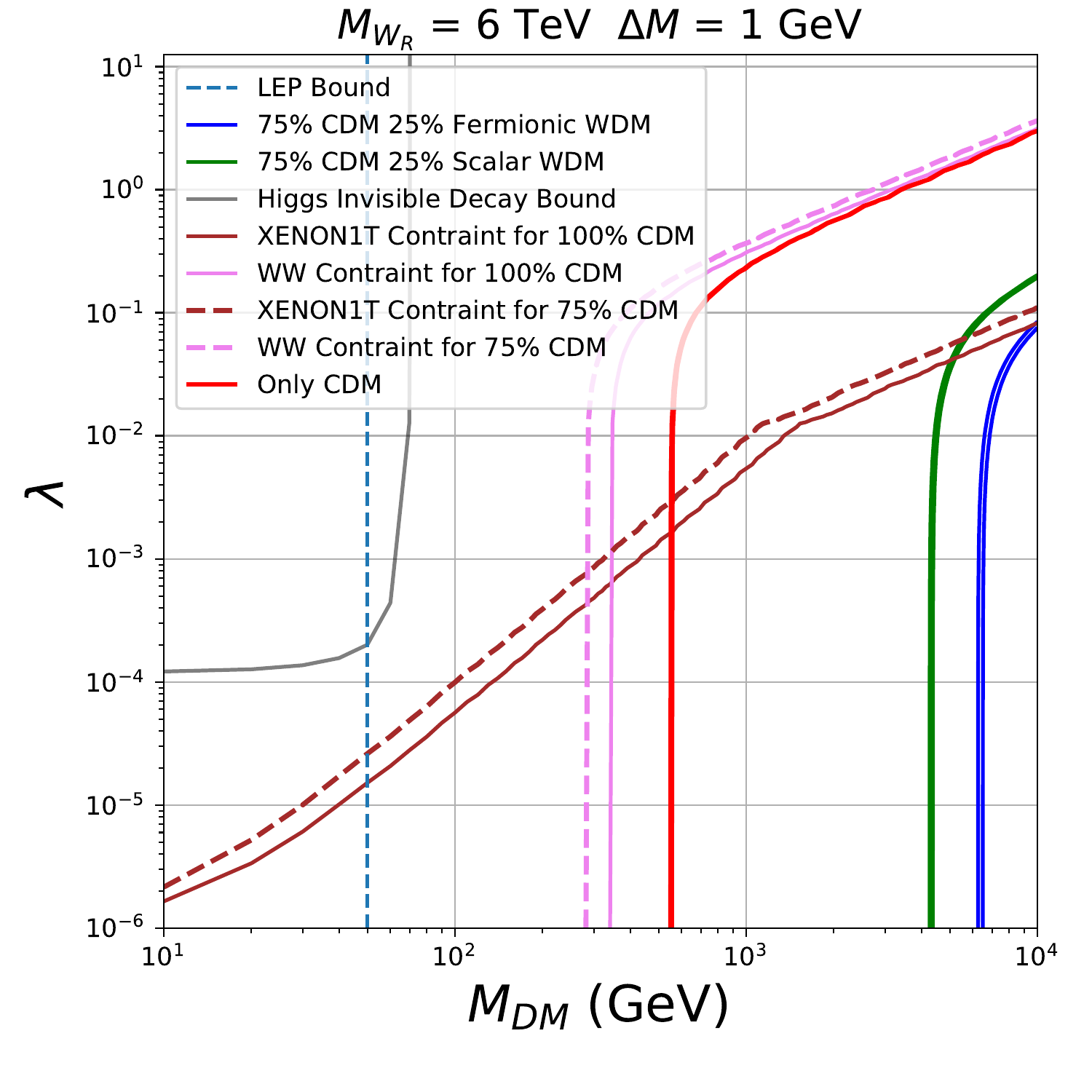,width=0.50\textwidth,clip=}
\epsfig{file=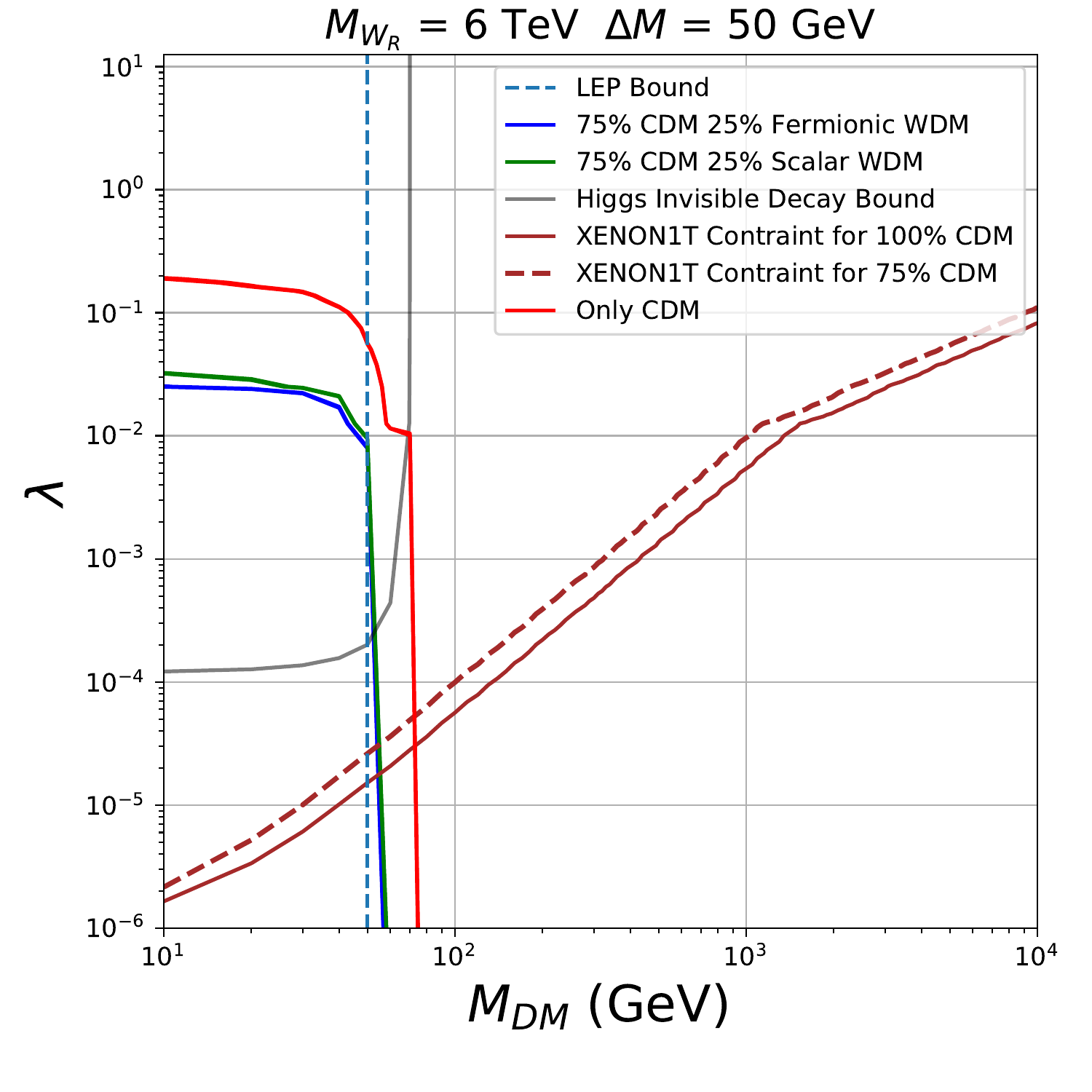,width=0.50\textwidth,clip=}
\end{tabular}
\caption{Allowed parameter space from relic abundance point of view for left scalar doublet dark matter assuming $100\%$ CDM (red solid line) and $75\%$ CDM (green solid line for scalar WDM, blue solid line for fermion WDM). The y-axis shows CDM-Higgs coupling and x-axis shows CDM mass. Choice of benchmark parameters and different relevant bounds are indicated by the legends.}
\label{fig2b}
\end{figure}

We show the parameter space for left scalar doublet DM for two different values of mass splitting within the components of the scalar doublet in figure \ref{fig2}, \ref{fig2a}, \ref{fig2b} for three different relative contributions of CDM and WDM. The results are somewhat different from the inert scalar doublet model discussed extensively in the literature \cite{Ma:2006km,Barbieri:2006dq,Cirelli:2005uq, LopezHonorez:2006gr, Honorez:2010re, LopezHonorez:2010tb, Borah:2012pu, Arhrib:2013ela, Dasgupta:2014hha, Borah:2017dqx, Borah:2017dfn}. For left scalar doublet, the DM relic abundance is primarily governed by three parameters: the Higgs-DM coupling $\lambda$, DM mass $M_{\eta^0_L} = M_{\rm DM}$ and mass splitting $\Delta M$ between different components of left scalar doublet $\eta_L$. For simplicity, we consider same mass splitting between lightest neutral scalar and the charged as well as neutral pseudoscalar components of the doublet. There are two mass regions which gives rise to correct relic abundance or a sizeable fraction of it, as seen from the figures. These correspond to the low mass regime below $W_L$ mass threshold and the high mass regime typically above 550 GeV. For a particular value of $\Delta M$, either both or one of these two regimes can be allowed from relic abundance criteria. The thermal DM abundance remains very much suppressed in the intermediate mass regime due to very large annihilations to electroweak gauge bosons. In the left panel plots of figure \ref{fig2}, \ref{fig2a}, \ref{fig2b} the low mass regime disappears as the DM coannihilations are very large due to smaller mass splitting $\Delta M$. The opposite happens in the right panel plots of these figures where the high mass regime disappears due to large mass splitting $\Delta M$. Although such behaviour is already known from previous studies, here we show the shift in allowed parameter space due to different relative abundance of left scalar doublet dark matter. Also, we show the change in CDM parameter space due to the change in $W_R$ mass in upper and lower panels of each of these figures \ref{fig2}, \ref{fig2a}, \ref{fig2b}. Although left scalar doublet annihilations are not affected by the $W_R$ boson, the WDM thermal abundance and its subsequent dilution depends crucially on $W_R$ mass and the late entropy dilution also affects the abundance of left scalar doublet DM.

Apart from relative relic abundance criteria, the next constraint we incorporate here are the ones from direct detection. Although the $Z_L$ boson mediated inelastic scattering of DM with nucleons is kinematically forbidden for the chosen mass splittings $\Delta M$, there can be tree level scattering processes of left scalar dark matter $\eta_L$ with nucleons mediated by the standard model Higgs. The relevant spin independent scattering cross section mediated by SM Higgs is given as \cite{Barbieri:2006dq}
\begin{equation}
 \sigma_{\text{SI}} = \frac{\lambda^2 f^2}{4\pi}\frac{\mu^2 m^2_n}{m^4_h m^2_{\eta_L}}
\label{sigma_dd}
\end{equation}
where $\mu = m_n m_{\eta_L}/(m_n+m_{\eta_L})$ is the $\eta_L$-nucleon reduced mass and $\lambda$ is the quartic coupling involved in $\eta_L$-Higgs interaction which was assumed to take specific values in the relic abundance plot shown in figures \ref{fig2}, \ref{fig2a}, \ref{fig2b}. A recent estimate of the Higgs-nucleon coupling $f$ gives $f = 0.32$ \cite{Giedt:2009mr} although the full range of allowed values is $f=0.26-0.63$ \cite{Mambrini:2011ik}. The latest bound on such spin independent DM-nucleon elastic scattering cross section comes from the one year exposure of the XENON1T experiment \cite{Aprile:2018dbl}. According to this, for a dark matter mass of 30 GeV, DM-nucleon scattering cross sections above $4.1 \times 10^{-47} \; \text{cm}^2$ are excluded at $90\%$ confidence level. This will constrain the $\eta_L$-Higgs coupling $\lambda$ significantly, if $\eta_L$ gives rise to most of the dark matter in the Universe. However, this bound gets relaxed a factor $n\%$ if $\eta_L$ corresponds to only $n\%$ of the total DM density. This is because the DM direct detection rate is directly proportional to the density of DM. One can also constrain the $\eta_L$-Higgs coupling $\lambda$ from the latest LHC constraint on the invisible decay width of the SM Higgs boson. This constraint is applicable only for dark matter mass $m_{\eta_L} < m_h/2$. The invisible decay width is given by
\begin{equation}
\Gamma (h \rightarrow \text{Invisible})= {\lambda^2 v^2\over 64 \pi m_h} 
\sqrt{1-4\,m^2_{\eta_L}/m^2_h}
\end{equation}
The latest ATLAS constraint on invisible Higgs decay is \cite{Aad:2015pla}
$$\text{BR} (h \rightarrow \text{Invisible}) = \frac{\Gamma (h \rightarrow \text{Invisible})}{\Gamma (h \rightarrow \text{Invisible}) + \Gamma (h \rightarrow \text{SM})} < 22 \%$$
Since this constraint is independent of astrophysical DM density, it remains same for all relative abundance of CDM considered in this work. We also incorporate the bound from LEP which restricts the sum of neutral scalar and pseudoscalar masses to be more than the $Z_L$ mass. Finally, we include constraints from indirect detection experiments coming from gamma ray searches that put upper bounds on DM annihilations to charged final states. For left scalar doublet, these bounds are more stringent for $W_L$ final states and hence we indicate the corresponding exclusion lines in figures \ref{fig2}, \ref{fig2a}, \ref{fig2b}. We discuss the details of indirect detection in an upcoming section. It is worth noting that indirect detection constraints get weaker by a factor $(n\%)^2$ if $\eta_L$ corresponds to only $n\%$ of the total DM density. This is due to the fact that DM annihilation rates are proportional to the DM density squared. 

\begin{figure}[!h]
\centering
\begin{tabular}{cc}
\epsfig{file=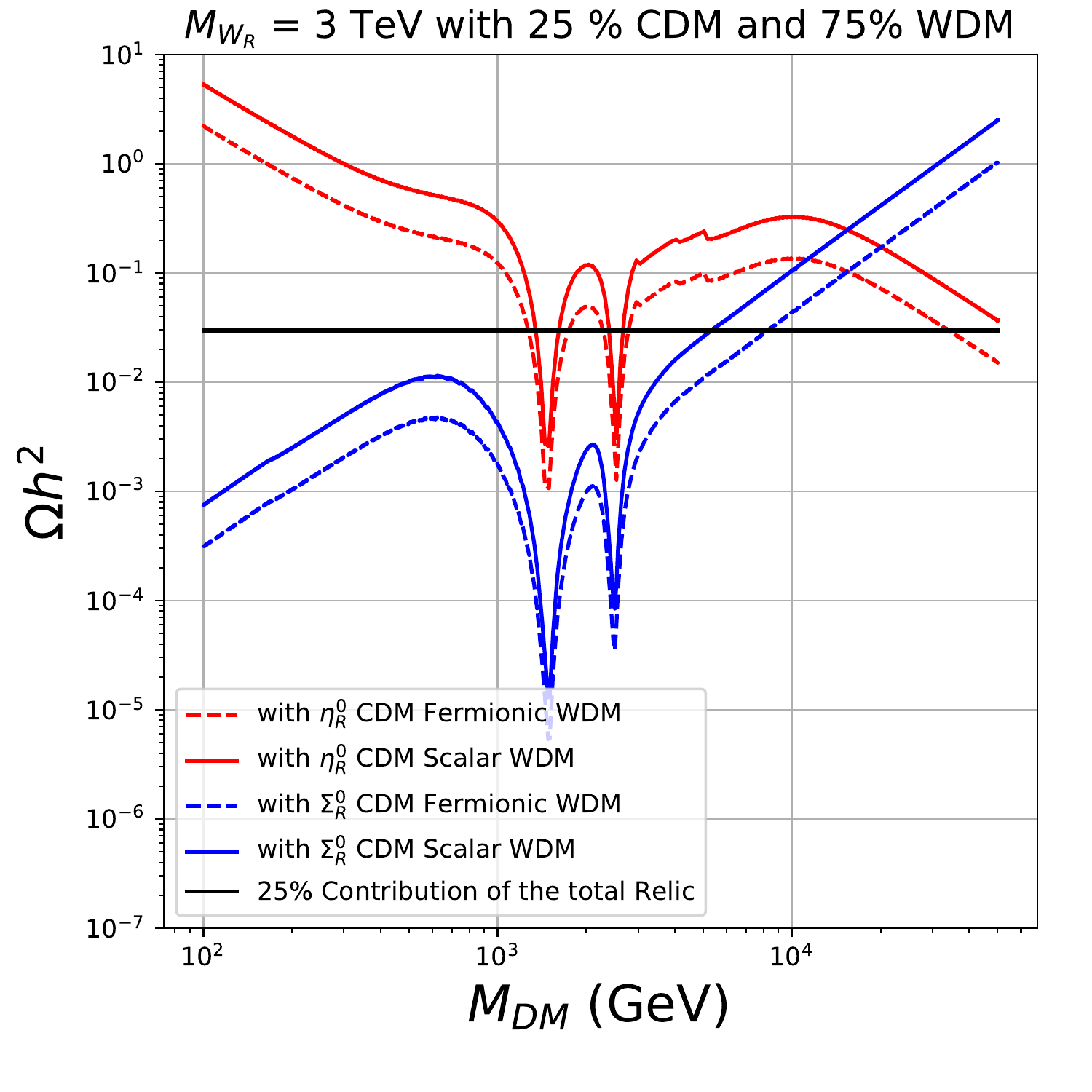,width=0.50\textwidth,clip=}
\epsfig{file=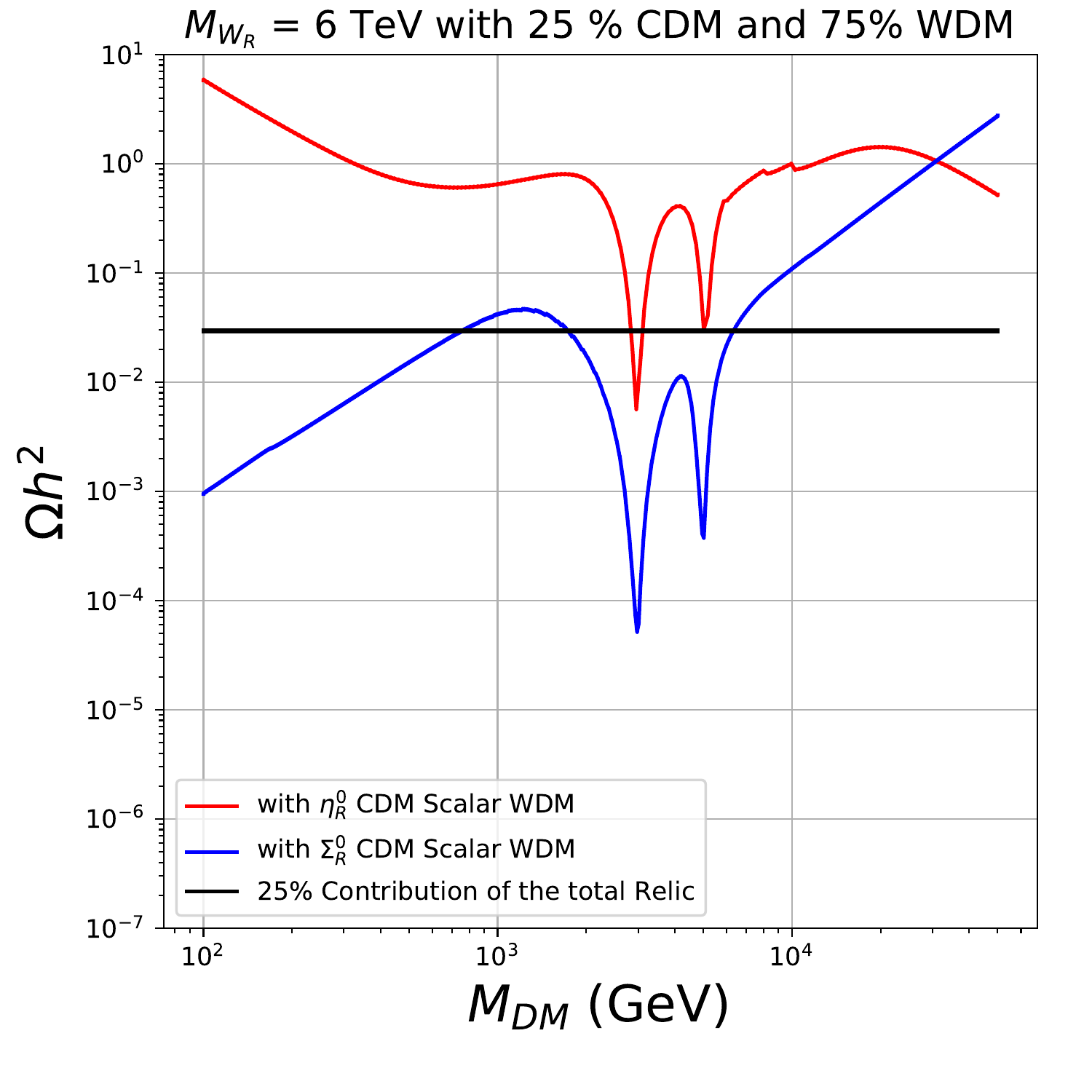,width=0.50\textwidth,clip=}
\end{tabular}
\caption{Relic abundance of right handed scalar (fermion) doublet (triplet) dark matter along with different relative contribution of fermion/scalar WDM.}
\label{fig3}
\end{figure}

\begin{figure}[!h]
\centering
\begin{tabular}{cc}
\epsfig{file=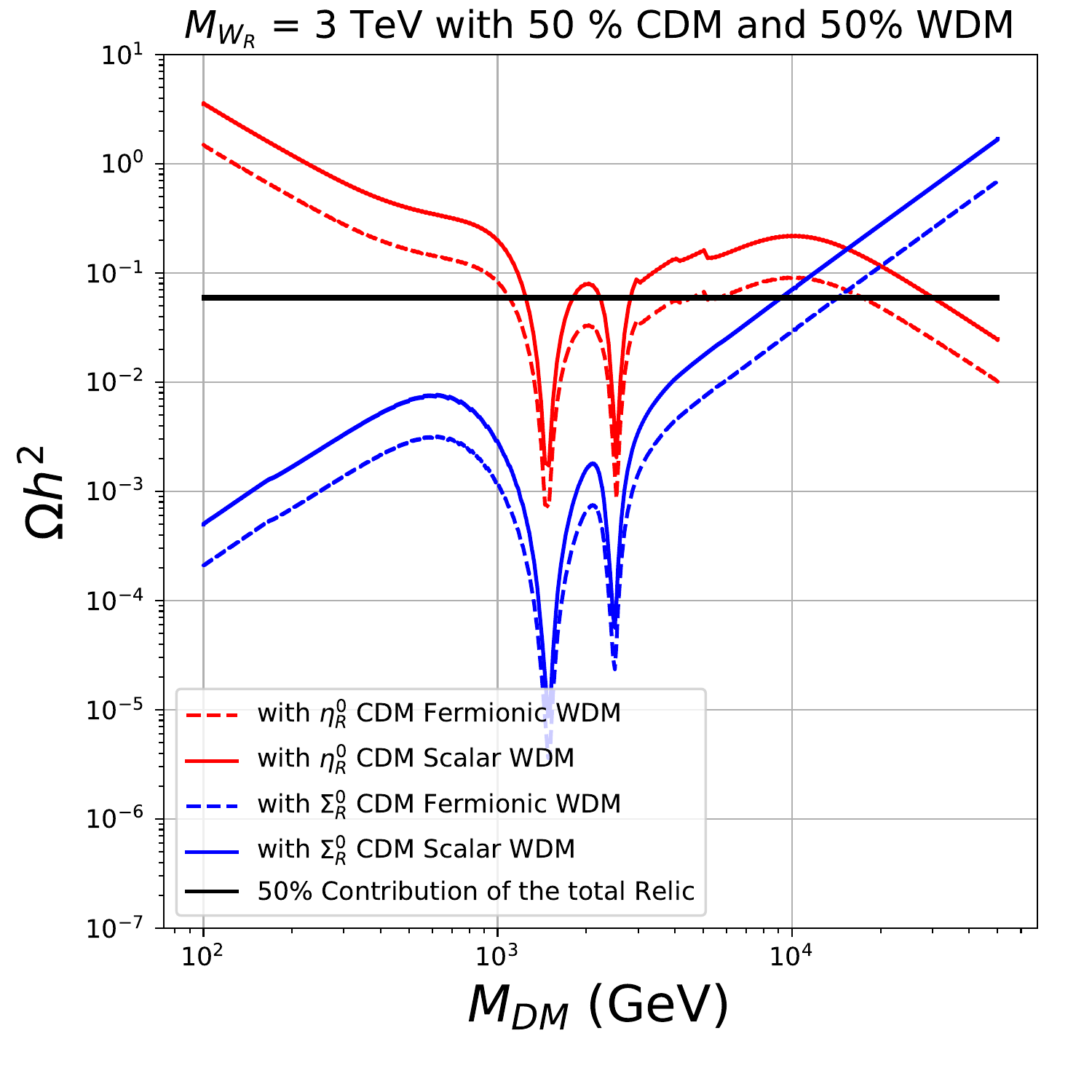,width=0.50\textwidth,clip=}
\epsfig{file=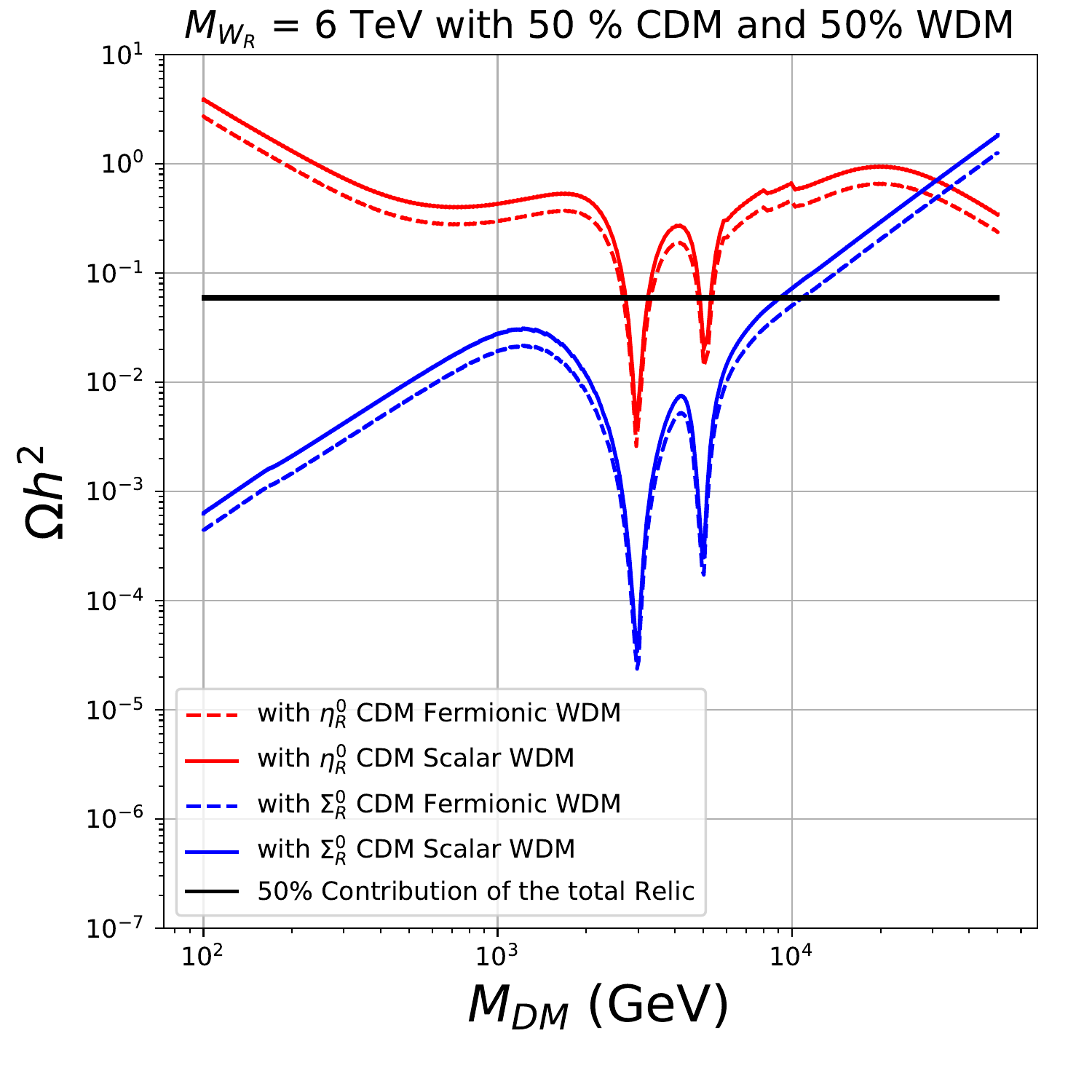,width=0.50\textwidth,clip=}
\end{tabular}
\caption{Relic abundance of right handed scalar (fermion) doublet (triplet) dark matter along with different relative contribution of fermion/scalar WDM.}
\label{fig3a}
\end{figure}

\begin{figure}[!h]
\centering
\begin{tabular}{cc}
\epsfig{file=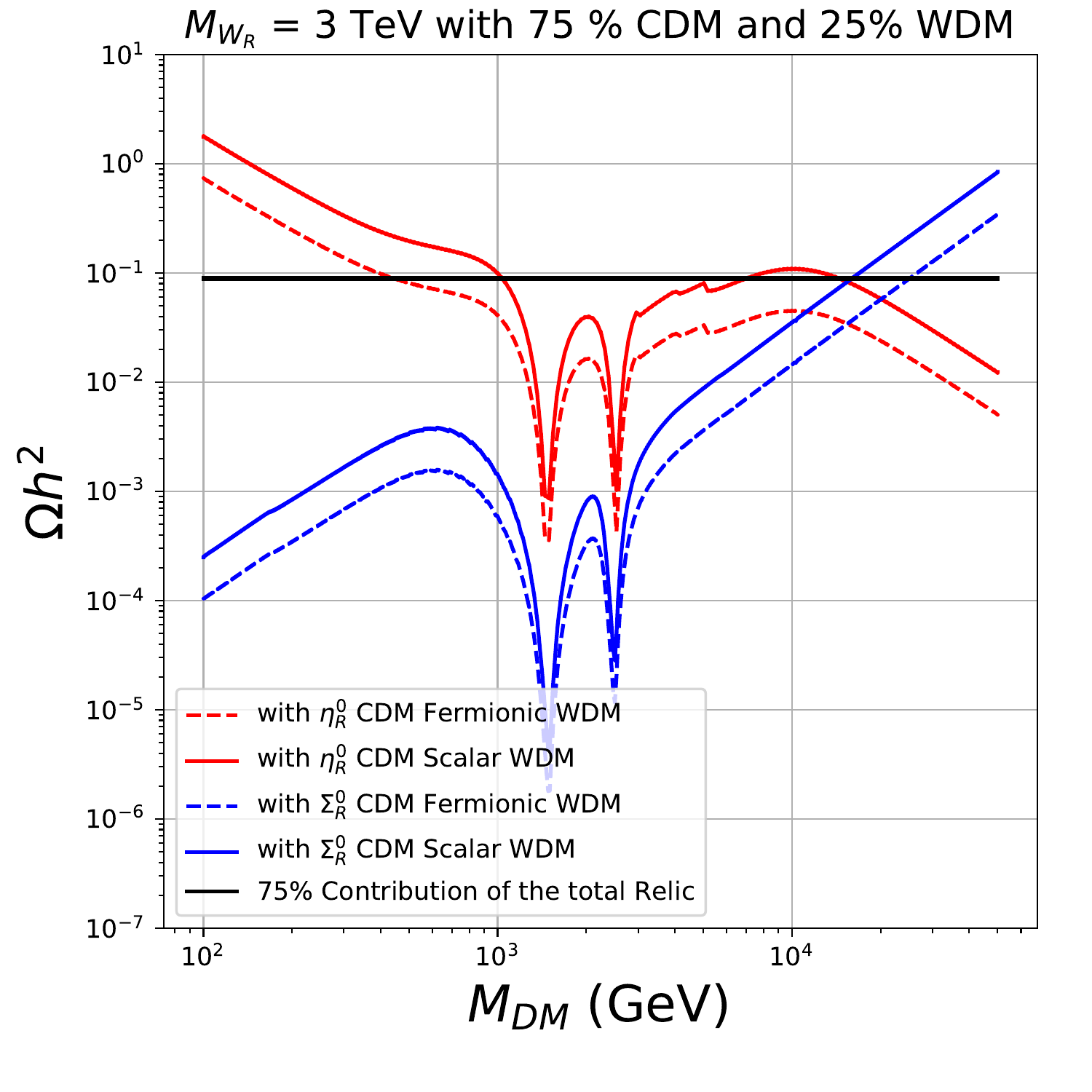,width=0.50\textwidth,clip=}
\epsfig{file=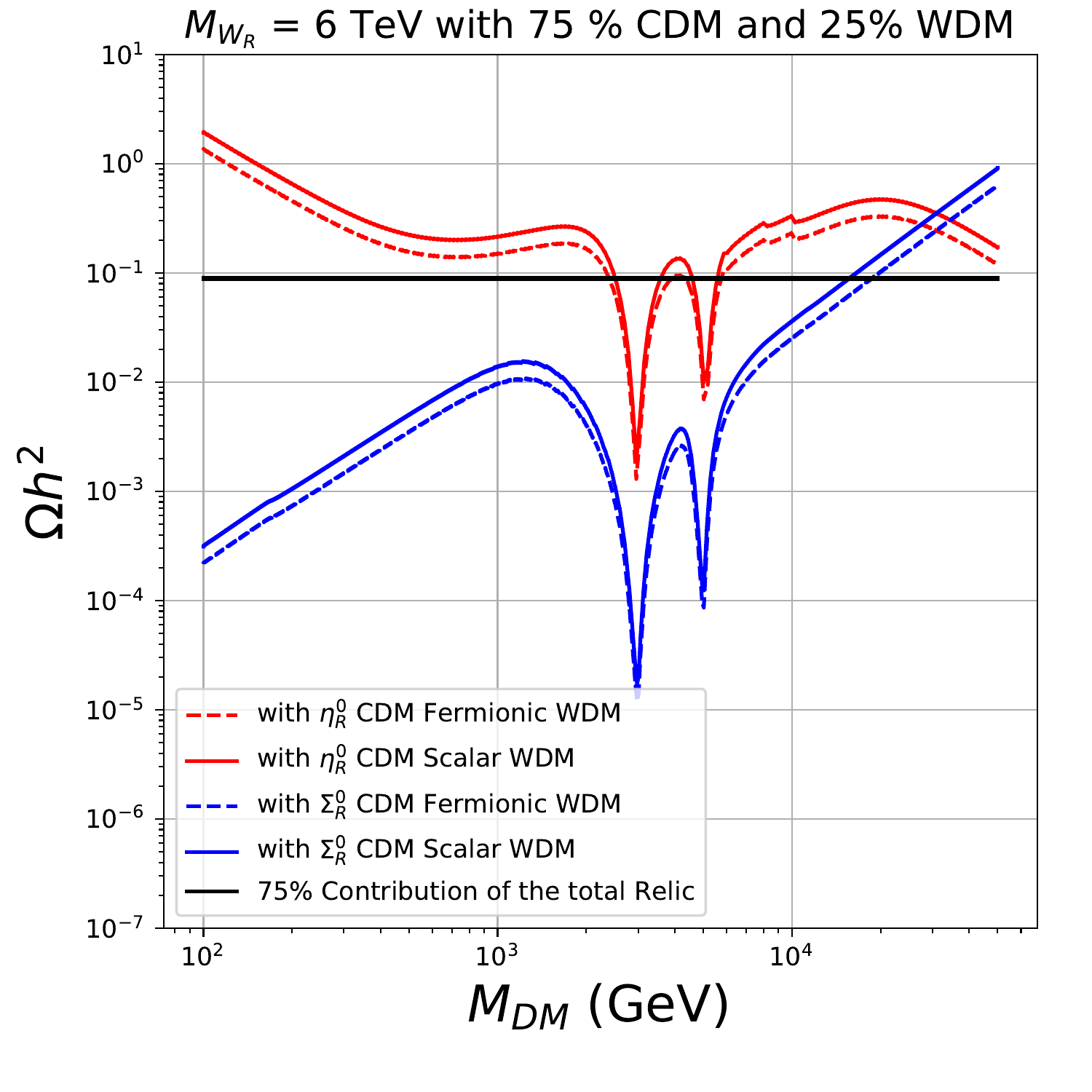,width=0.50\textwidth,clip=}
\end{tabular}
\caption{Relic abundance of right handed scalar (fermion) doublet (triplet) dark matter along with different relative contribution of fermion/scalar WDM.}
\label{fig4}
\end{figure}

We show the parameter space giving rise to different fraction of total DM abundance for right scalar doublet DM as well as right fermion triplet DM in figure \ref{fig3}, \ref{fig3a}, \ref{fig4} for different relative contributions of CDM and WDM. The results can be compared with the ones shown in \cite{Garcia-Cely:2015quu, Borah:2016hqn, Borah:2017leo}, but extended here for different fractions of total DM abundance. Similar to these works, here also we do not take the Higgs portal interactions of right scalar doublet DM so that the DM relic abundance is more sensitive to the $SU(2)_R$ gauge sector. In this approximation, the DM relic abundance is mainly governed by the DM coannihilations through $W_R, Z_R$ bosons. For the right handed scalar doublet, the mass splitting between DM and other components of the doublet is taken to be 50 GeV.

Since the fermion triplet can annihilate only through gauge interactions, the number of free parameters affecting the relic abundance is less, in fact only two namely, the DM mass and the right handed gauge boson masses (also the $SU(2)_R$ gauge coupling $g_R$ which is taken to be same as $g_L$ in the left-right symmetric limit). Unlike scalar DM, here the mass splitting between charged and neutral components of the fermion triplet $\Sigma_R$ is not a free parameter but generated at one loop level through gauge boson corrections. This mass splittings between charged and neutral components of right-handed triplet fermion is given by
\begin{align}
M_{\Sigma_R^\pm}-M_{\Sigma_R^0} &\simeq \frac{\alpha_2}{4\pi} \frac{g_R^2}{g_L^2} M \left[ f(r_{W_R}) - c_M^2 f(r_{Z_R})-s_W^2 s_M^2 f(r_{Z_L})- c_W^2 s_M^2 f(r_\gamma)\right] .
\label{eq:RH_mass_splitting}
\end{align}
Here the one loop self-energy corrections through mediations of gauge bosons are presented within the square bracket. For example,  the mass splitting with the approximation $M_\Sigma \gg M_{W_R}$ goes as $\alpha_{2} \left( M_{W_R} - c_M^2 M_{Z_R}\right)/2$. The sine and cosine of different angles $c_M, c_W, s_M, s_W$ etc. correspond to the angles involved in the rotation of neutral gauge bosons given by
\begin{equation}
\left(\begin{array}{c}W_{L \mu}^3\\W_{R \mu}^3\\B_\mu\end{array}\right)=
\left(\begin{array}{ccc}
c_{W}c_{\phi}    &  c_W s_{\phi}                           &  s_{W} \\
-s_{W} s_M c_{\phi}-c_M s_{\phi}       & -s_W s_M s_{\phi}+c_M c_{\phi}   & c_{W} s_M    \\
-s_W c_M c_{\phi}+s_M s_{\phi}   & -s_W c_M s_{\phi}-s_M c_{\phi}   &  c_W c_M
\end{array} \right)
\left(\begin{array}{c}Z_{L \mu}\\Z_{R \mu}\\A_\mu\end{array}\right). 
\label{neutral1}
\end{equation}
Also, $r_B = \frac{M_B}{M_{\Sigma}}$ and the loop function $f(r)$ being given as
\begin{equation}
f(r) \equiv 2 \int^1_0 dt (1+t) \ln\left[t^2+(1-t)r^2 \right]
\label{eq:fr}
\end{equation}
Since the mass splitting decides the amount of coannihilations, choosing the mass of right handed gauge bosons and DM mass is enough to predict the thermal abundance of DM shown in figures \ref{fig3}, \ref{fig3a}, \ref{fig4}. Similar to left scalar doublet case, here also we consider two different values of $W_R$ mass namely 3 TeV and 6 TeV and the corresponding results are shown in left and right panels respectively of figures \ref{fig3}, \ref{fig3a}, \ref{fig4}.

One can similarly calculate the abundance of left fermion triplet DM as well for which the mass splitting due to electroweak gauge corrections is 
\begin{align}
\begin{split}
M_{\Sigma_L^\pm}-M_{\Sigma_L^0} &\simeq  \alpha_2  M_{W} \sin^2 (\theta_W/2) + \mathcal{O}(M_{W}^3/M^2_\Sigma).
\end{split}
\end{align}
Since there is only one free parameter that decides fermion triplet abundance which is its mass, we do not have any parameter space to show for it. As shown first by the authors of \cite{Ma:2008cu}, such a triplet satisfies correct relic abundance only for DM mass around 3 TeV. It is important to emphasise that, for such heavy fermion triplet DM, Sommerfeld effects \cite{Hisano:2003ec,Hisano:2004ds, Hisano:2006nn} are important as the corresponding gauge bosons masses are very small compared to DM mass. Including such effects, pushes the allowed DM mass slightly beyond 3 TeV \cite{Hisano:2006nn}. Such effects are not important in the right fermion triplet DM, as long as the triplet mass is comparable to right handed gauge boson masses. Detailed calculation of such effects is beyond the scope of the present work and can be found in above references and also in \cite{Garcia-Cely:2015quu} within the context of LRSM.

Since the CDM freeze-out occurs around $T_F \approx M_{\rm CDM}/x_F, x_F \approx 20-30$, the entropy dilution near the MeV temperature affects their freeze-out abundance too. This along with the fact that we demand less than $100\%$ contribution from CDM to the total DM density, give rise to different parameter space compared to the ones studied in the literature where thermal abundance of CDM is constrained to be the exact DM abundance. We find it very interesting because, even if some part of DM parameter space remains disallowed due to insufficient annihilation cross section leading to overproduction, such a multi-component scenario will allow it due to the late time entropy release. This can have interesting implications for heavy DM scenarios where DM typically gets overproduced due to an upper bound on annihilation cross section from unitarity limit \cite{Nussinov:2014qva}. As the reference \cite{Nussinov:2014qva} mentions, this unitarity limit in fact rules out DM masses beyond a few tens of TeV making it difficult to construct heavy DM mass models which can otherwise be interesting from IceCube experiment point of view. Recently a PeV DM scenario within LRSM was discussed in \cite{Borah:2017xgm}, as a possible explanation of the PeV neutrino events at IceCube, where late time entropy dilution was invoked to bring the DM abundance within limits. In the present scenario however, such entropy release plays another non-trivial role in reducing the WDM abundance.

\begin{figure}[!h]
\centering
\epsfig{file=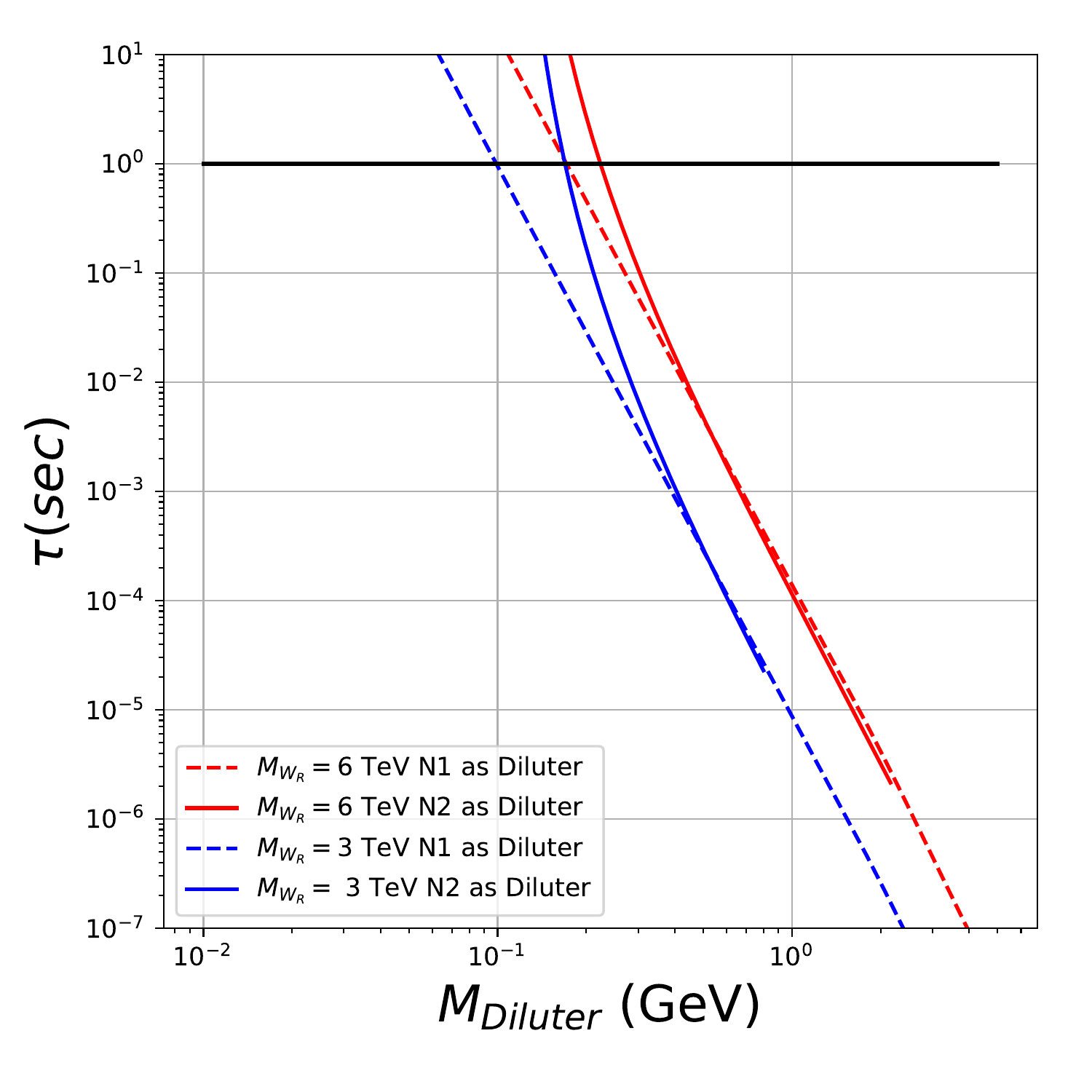,width=0.750\textwidth,clip=}
\caption{Lifetime of right handed neutrinos including only the gauge boson mediated decay channels.}
\label{fig61}
\end{figure}

\subsection{Relic Abundance of WDM}

After showing the parameter space giving rise to the desired relic abundance of CDM, we move on to discussing the requirements for keV scale WDM. For fermion WDM that is $N_1$, the final relic abundance after entropy dilution due to $N_2$ decay discussed above is given by 
\begin{equation}
\frac{\Omega_{N_1}}{\Omega_{\text{DM}}} \approx 1.0 \left( \frac{107}{g_{*f}} \right) \left( \frac{M_{N_1}}{ 1 \;\text{keV}} \right) \left ( \frac{ 1\; \text{sec}} {\tau_{N_2}} \right)^{1/2} \left ( \frac{1 \; \text{GeV}}{ M_{N_2}} \right) \left ( \frac{g_*(T_{fN_2})}{60} \right)
\end{equation}
Similarly, the abundance of scalar WDM after entropy dilution due to the decay of $N_2$ is
\begin{equation}
\frac{\Omega_{\Omega^0_R}}{\Omega_{\text{DM}}} \approx 1.0 \left( \frac{107}{g_{*f}} \right) \left( \frac{M_{\Omega^0_R}}{ 1 \;\text{keV}} \right) \left ( \frac{ 1\; \text{sec}} {\tau_{N_1}} \right)^{1/2} \left ( \frac{1 \; \text{GeV}}{ M_{N_1}} \right) \left ( \frac{g_*(T_{fN_1})}{84} \right)
\end{equation}
It can be seen from figure \ref{fig61} that the required lifetime of decaying particles releasing entropy can be achieved for suitable values of right handed gauge boson masses, taking into account of gauge mediated decay channels only. If we consider non-zero and large left-right neutrino mixing $\theta_{\nu}$, the decay lifetime can be even shorter, insufficient for the correct entropy dilution. The heavy-light neutrino mixing $\theta_{\nu}$ can be kept small in left-right models, by appropriate tuning of Yukawa couplings. This can at the same time be consistent with correct neutrino mass due to the existence of additional contribution (type II seesaw).

\subsection{Indirect Detection}
Due to the existence of two DM components with widely separated mass scales, the models discussed in this work can have very interesting indirect detection signatures different from single component DM. Since the CDM has mass in the GeV-TeV scale and WDM has keV scale mass in this setup, they can annihilate or decay into SM particles with different energies. Among such SM final state particles, photons and neutrinos, being electromagnetically neutral, have the potential to reach the indirect detection experiments without getting affected in the intermediate regions. If DM is of CDM type with typical masses in the GeV-TeV scale, such photons lie in the gamma ray regime whereas for keV scale WDM they correspond to X-ray part of the electromagnetic spectrum.
\begin{figure}[!h]
\centering
\begin{tabular}{cc}
\epsfig{file=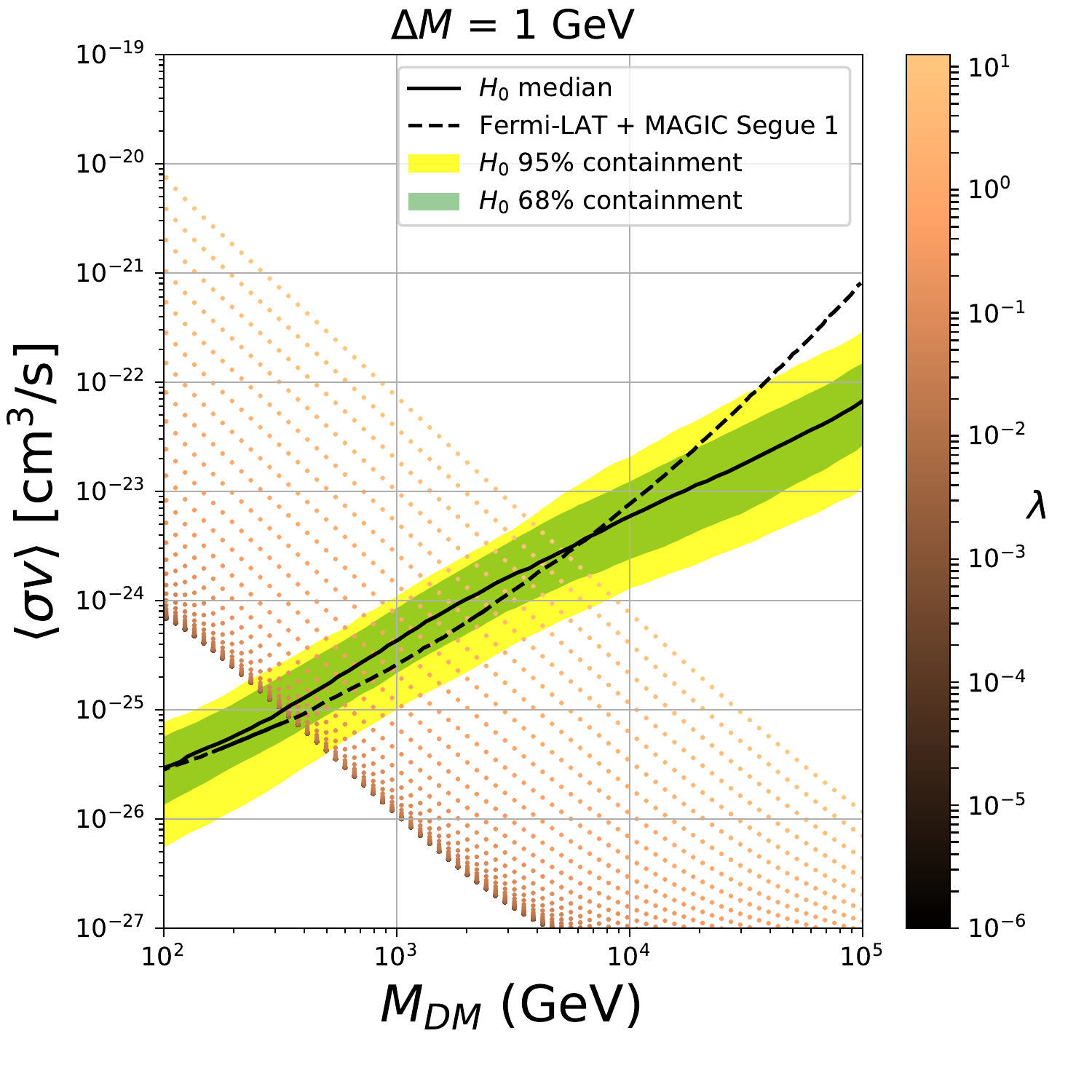,width=0.50\textwidth,clip=}
\epsfig{file=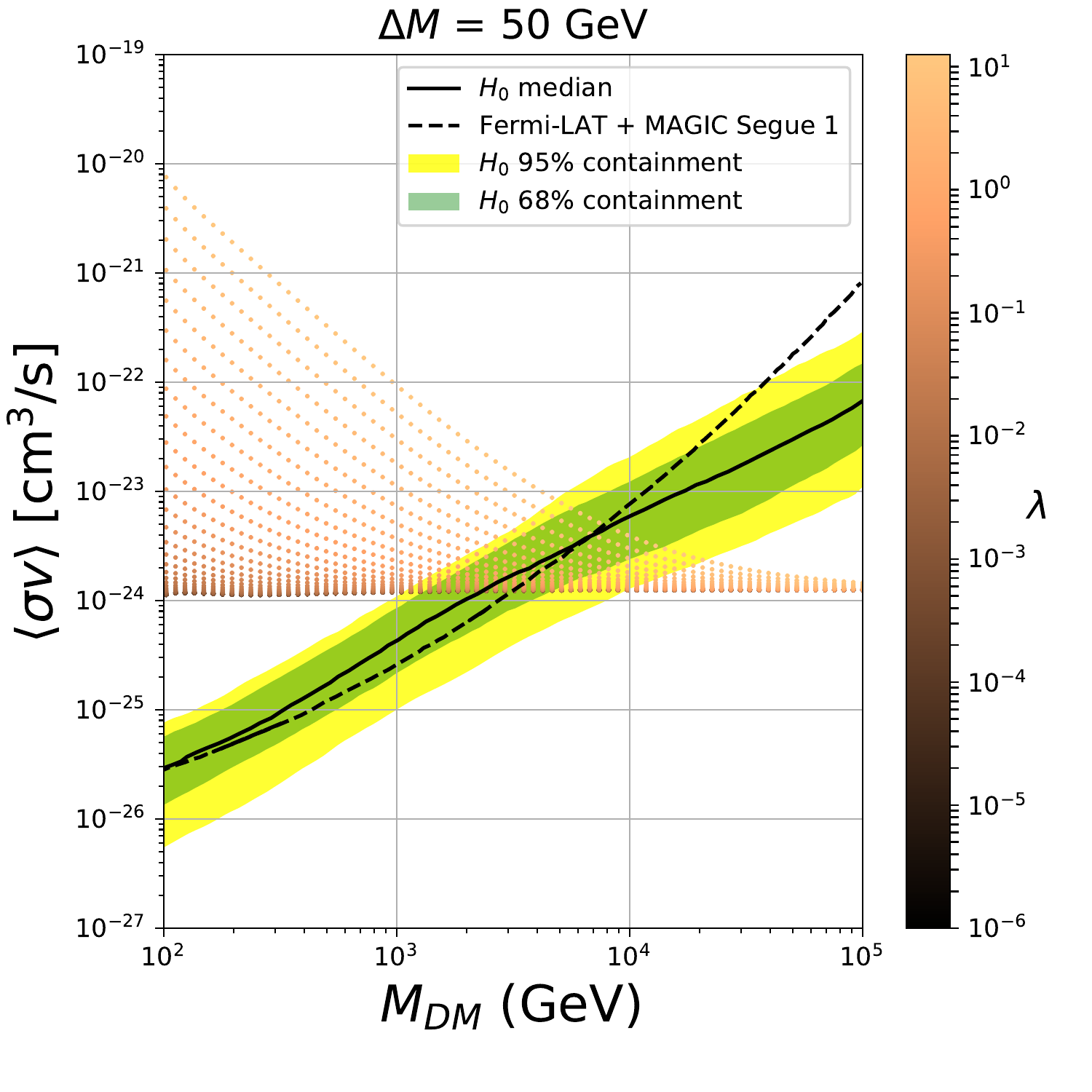,width=0.50\textwidth,clip=}
\end{tabular}
\caption{Left scalar doublet DM annihilation into $W^+ W^-$ final states for mass splitting 1 GeV and 50 GeV respectively. Thick solid lines show the limits obtained by combining Fermi-LAT observations of 15 dSphs with MAGIC observations of Segue 1. The thin-dotted line, green and yellow bands show,  respectively,  the median and the symmetrical, two-sided $68\%$ and $95\%$ containment bands for the distribution of limits under the null hypothesis, details of which can be found in \cite{Ahnen:2016qkx}.}
\label{fig6a}
\end{figure}
The observed differential gamma ray flux produced due to DM annihilations is given by
\begin{equation}
\frac{d\Phi}{dE} (\triangle \Omega) = \frac{1}{4\pi} \langle \sigma v \rangle \frac{J (\triangle \Omega)}{2M^2_{\text{DM}}} \frac{dN}{dE}
\end{equation}
where $\triangle \Omega$ is the solid angle corresponding to the observed region of the sky, $\langle \sigma v \rangle$ is the thermally averaged DM annihilation cross section, $dN/dE$ is the average gamma ray spectrum per annihilation process and the astrophysical $J$ factor is given by
\begin{equation}
J(\triangle \Omega) = \int_{\triangle \Omega} d\Omega' \int_{LOS} dl \rho^2(l, \Omega').
\end{equation}
In the above expression, $\rho$ is the DM density and LOS corresponds to line of sight. Therefore, measuring the gamma ray flux and using the standard astrophysical inputs, one can constrain the CDM annihilation into different charged final states like $\mu^+ \mu^-, \tau^+ \tau^-, W^+_L W^-_L, b\bar{b}$. Since DM can not couple to photons directly, gamma rays can be produced from such charged final states. Using the bounds on DM annihilation to these final states arising from the global analysis of the Fermi-LAT and MAGIC observations of dSphs \cite{Ahnen:2016qkx}, we show the status of our model for different benchmark values of parameters. It turns out that some region of parameter space for left scalar doublet CDM studied in this work, the indirect detection constraints are rather severe, if CDM constitutes $100\%$ of the DM of the Universe. Here we see, how a multi-component DM scenario can help us to relax such bounds.

Since these bounds are more severe for low mass DM, we apply it to left scalar doublet DM where there is a possibility to get the desired relic in the low mass regime, unlike in the case of right handed (scalar as well as fermion) dark matter. We first show the CDM (left scalar doublet) annihilations into $W^+_L W^-_L$ final states for two different values of DM-Higgs coupling $\lambda_L$ in figure \ref{fig6a}. It can seen that, as we increase the mass splitting between the components of the scalar doublet, the annihilation cross section to $W^+_L W^-_L$ final states increases, specially in the high mass regime. In fact, we find that the mass splitting of 50 GeV is ruled out by the Fermi-LAT+MAGIC bounds even beyond DM mass of 1 TeV. While these plots show the overall behaviour of the annihilation cross section as a function of mass, it does not incorporate the constraints from relic abundance criteria. For a more realistic comparison of these bounds along with the relic constraints, please refer to the figures \ref{fig2}, ref{fig2a}, \ref{fig2b} discussed earlier. Similarly, we can compare the DM annihilation to charged fermion final states like muon, tau, bottom quarks with the corresponding limits. These limits are severe only near the Higgs resonance $M_{\rm DM} \approx m_h/2$ where the thermal abundance of CDM is anyway suppressed.

\begin{figure}[!h]
\centering
\epsfig{file=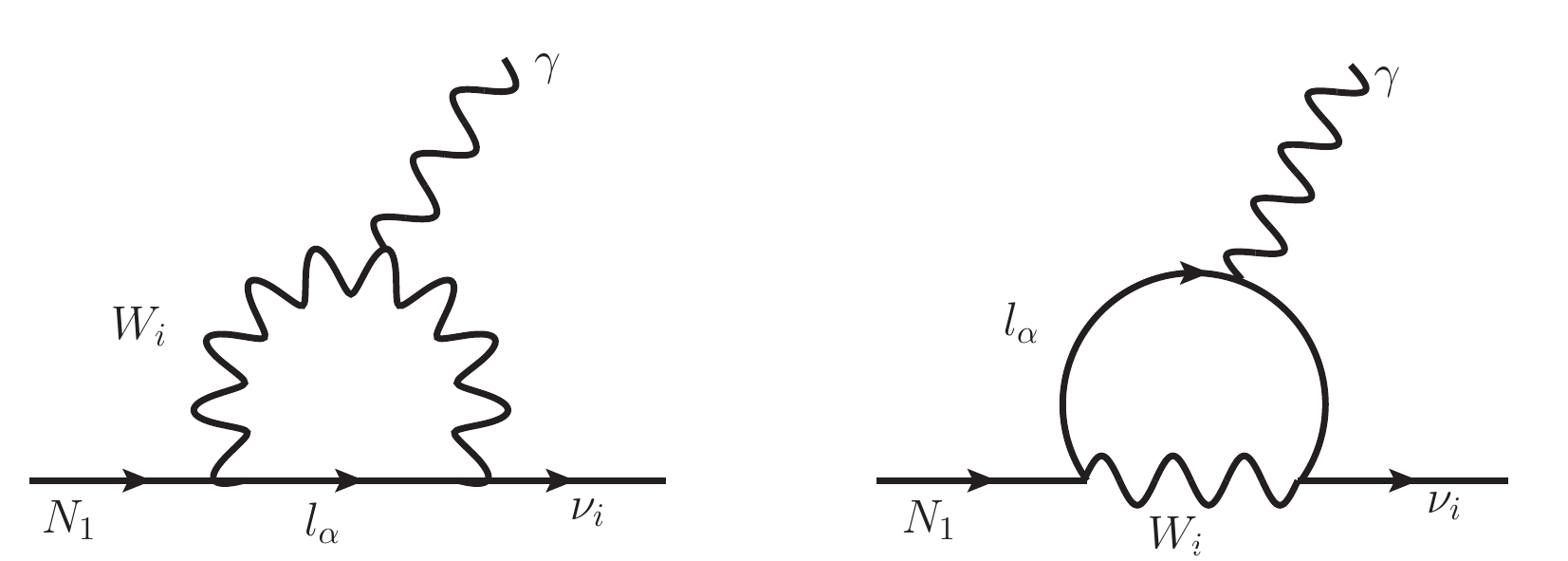,width=0.75\textwidth,clip=}
\caption{Heavy neutrino decay into a light one and a photon at one loop.}
\label{fig7}
\end{figure}

For keV scale WDM, one can have another interesting signature at indirect detection experiments in terms of a monochromatic X-ray line. Such a monochromatic X-ray line could have already been seen by the XMM-Newton X-ray telescope as mentioned earlier. Though one requires future data to confirm this claim as discovery, it nevertheless motivates one to look for rich particle physics explanations. In the present models, such a decay can occur due to the long lived nature of both fermion and scalar WDM. The fermion WDM, which is the lightest right handed neutrino, can decay into a photon and a light neutrino at one loop level through the diagrams shown in figure \ref{fig7}. Since the light neutrinos can be considered to be almost massless, such a decay can lead to the final states carrying energy $M_{WDM}/2$ each. The two processes arise due to heavy-light neutrino mixing and $W_L-W_R$ mixing. The one loop $W_L-W_R$ mixing is given in \eqref{WLWRmixing} and heavy-light neutrino mixing was given in \eqref{hlmixing1}. In figure \ref{fig7}, the heavy and light neutrino mass eigenstates are denoted by $N_1, \nu_i$ respectively whereas $W_{i}, (i=1,2)$ correspond to the physical mass eigenstates of $W_L, W_R$ gauge bosons. The decay width is given as 
\begin{equation}
\Gamma_{N_1 \rightarrow \gamma \nu} = \frac{(m^2_{N_1} - m^2_\nu)^3}{16\pi m^3_{N_1}}\left(|\sigma_L|^2 + |\sigma_R|^2\right)
\end{equation}
where 
\begin{align}
\sigma_L &= ie g^2\left[-\frac{1}{2} \sin^2(\text{$\theta_{LR} $}) \sin (2 \theta_\nu )g(m_{W_1},t_{W_1}) - \frac{1}{2} \cos ^2(\theta_{LR} ) \sin (2 \theta_\nu )g(m_{W_2},t_{W_2}) \right. \nonumber \\
&+ \left.  \frac{1}{2} \sin (2 \theta_{LR}) \sin ^2(\theta_\nu )h(m_{W_1},t_{W_1}) - \frac{1}{2} \sin (2 \theta_{LR}) \sin ^2(\theta_\nu )h(m_{W_2},t_{W_2})\right] \\
\sigma_R &= ie g^2\left[\frac{1}{2} \sin^2(\text{$\theta_{LR} $}) \sin (2 \theta_\nu )g(m_{W_2},t_{W_2}) + \frac{1}{2} \cos ^2(\theta_{LR} ) \sin (2 \theta_\nu )g(m_{W_1},t_{W_1}) \right. \nonumber \\
&- \left.  \frac{1}{2} \sin (2 \theta_{LR}) \cos ^2(\theta_\nu )h(m_{W_1},t_{W_1}) + \frac{1}{2} \sin (2 \theta_{LR}) \cos ^2(\theta_\nu )h(m_{W_2},t_{W_2})\right] \\
\end{align}
\begin{align}
g(mB,t_B) &= \frac{ m_{N_1}}{64 \pi ^2 mB^4 \left(t_B-1\right)^4} \left(-3 mB^2 \left(t_B-1\right)^2 \left(\left(t_B-5\right) t_B+2\right) \right. \nonumber \\
&- \left.2 mB^2 \left(t_B \left(t_B \left(3
   t_B-5\right)+16\right)-8\right) \log \left(t_B\right)\right)\\
h(mB,t_B) &= -\frac{ \sqrt{t_B}}{64 \pi ^2 mB^3 \left(t_B-1\right)^4} \left(\left(t_B-1\right)^2 \left(\left(t_B+1\right) m_{N_1}^2+8 mB^2 \left(t_B \left(t_B+2\right)-2\right)\right) \right. \nonumber \\
&- \left. 2 \log
   \left(t_B\right) \left(t_B \left(t_B+1\right) m_{N_1}^2+2 mB^2 \left(t_B-1\right) \left(\left(t_B-1\right) t_B \left(2
   t_B+5\right)+8\right)\right)\right) 
\end{align}
and $t_B = (m_l/m_B)^2$.

\begin{figure}[!h]
\centering
\begin{tabular}{cc}
\epsfig{file=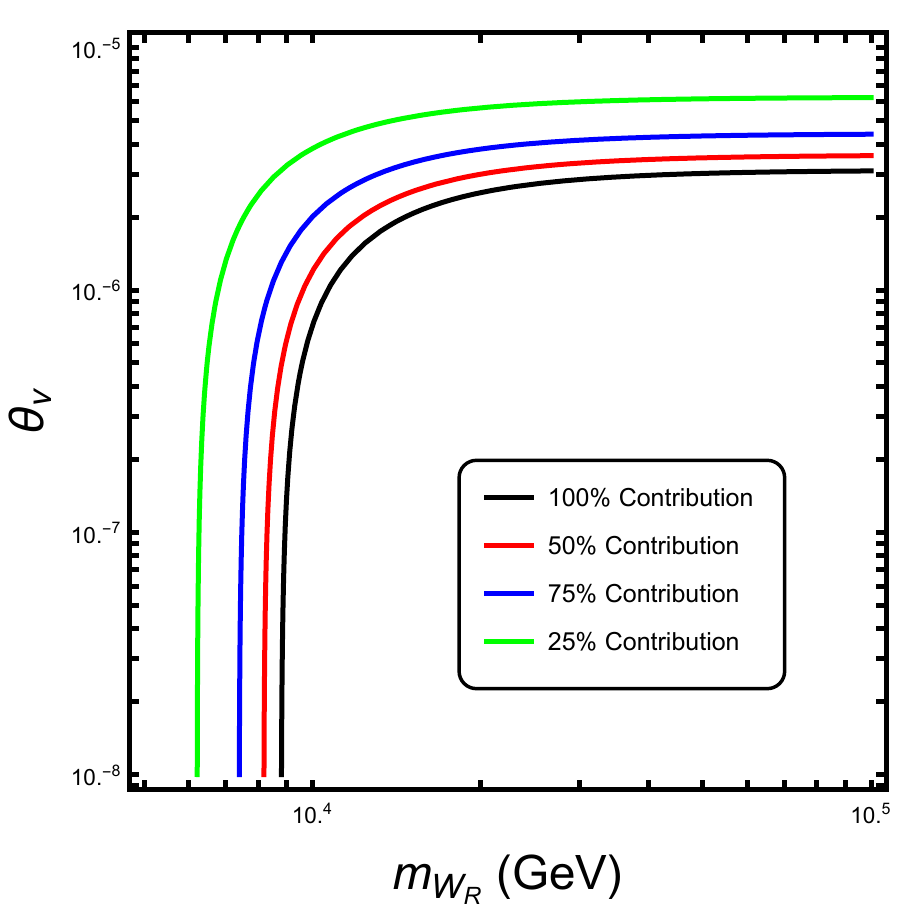,width=0.50\textwidth,clip=}
\epsfig{file=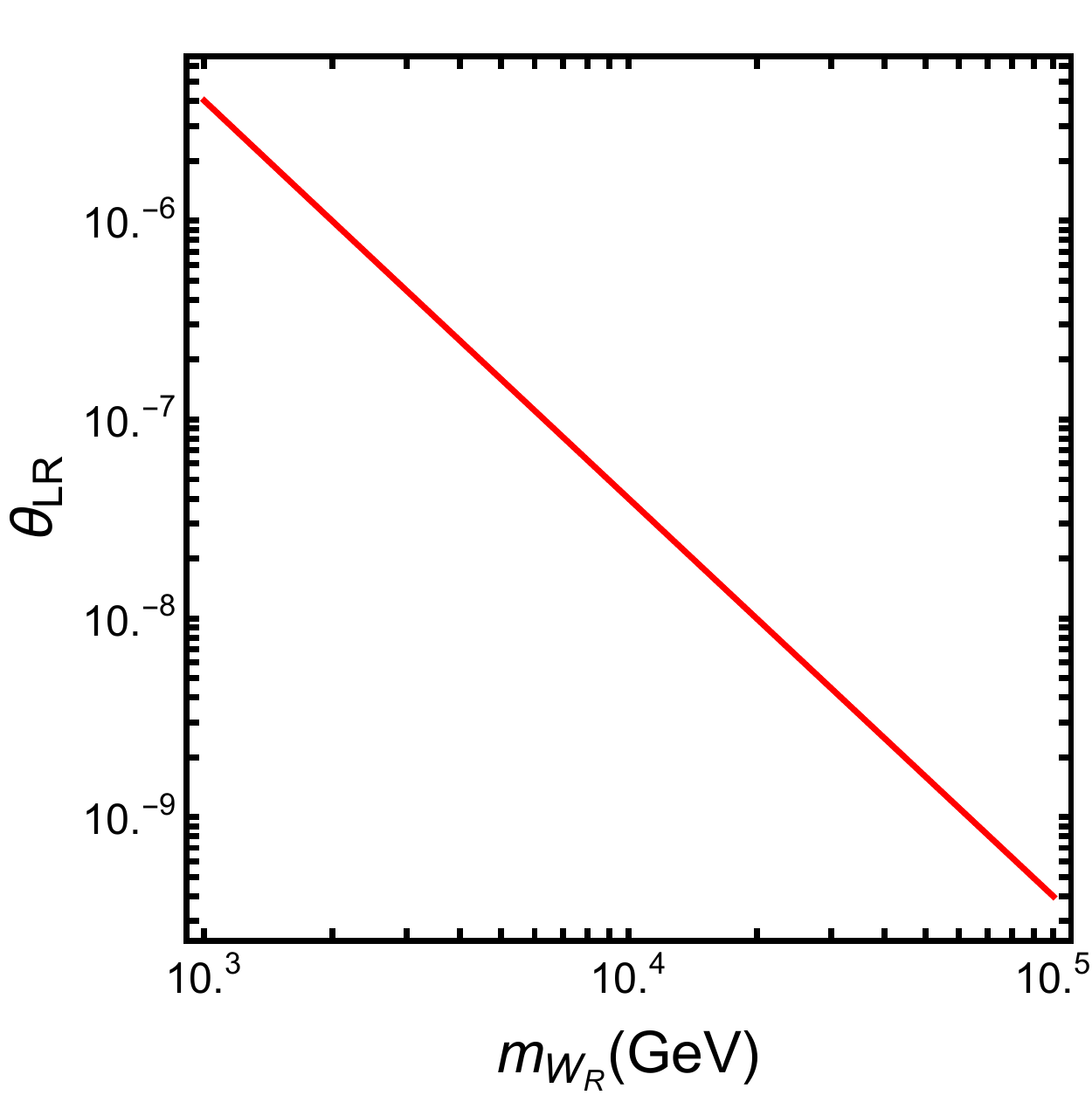,width=0.50\textwidth,clip=}
\end{tabular}
\caption{Left panel: Constraint on heavy-light neutrino mixing $\theta_{\nu}$ and $W_R$ mass from the requirement of producing the correct decay width of 7.1 keV sterile neutrino dark matter, producing the 3.55 keV X-ray line. Right panel: Left-right mixing $\theta_{LR}$ generated at one loop, as a function of $W_R$ mass.}
\label{fig7a}
\end{figure}

To give rise to the 3.55 keV line \cite{Bulbul:2014sua, Boyarsky:2014jta, Boyarsky:2014ska} from the decay of a 7.1 keV DM particle, one requires a lifetime of the order of $10^{28}$ s. For lightest right handed neutrino dark matter, such a lifetime can be generated for the mixing angles $\theta_{LR}, \theta_{\nu}$ shown in the plots of figure \ref{fig7a}. Since $\theta_{LR}$ depends upon $W_R$ mass as shown by the right panel of figure \ref{fig7a} as well as equation \eqref{WLWRmixing}, it is enough to vary $W_R$ mass and heavy-light neutrino mixing $\theta_{\nu}$ so that the correct lifetime of the keV DM is obtained. This is shown in the left panel plot of figure \ref{fig7a}. For $W_R$ mass below 10 TeV, it can be seen that the correct lifetime can be obtained for any value of $\theta_{\nu}$. This is due to the fact that for such $W_R$, the other mixing angle $\theta_{LR}$ can be sizeable and give the correct lifetime. However, if the $W_R$ mass is increased further, the left-right mixing $\theta_{LR}$ falls sharply and hence the heavy-light neutrino mixing $\theta_{\nu}$ has to dominate the decay process requiring it to be larger, as seen from the left panel plot of figure \ref{fig7a}. The fact that the value of $\theta_{\nu}$ does not change in the high mass regime of $W_R$ is due to the fact that for such values of heavy light neutrino mixing, the $W_L$ mediated diagram dominates and hence the decay width is almost independent of $W_R$ mass. As the relative abundance of WDM is decreased, the allowed parameter space in $\theta_{\nu}-m_{W_R}$ plot shifts towards left and upward, as seen from the left panel plot of figure \ref{fig7a}. This is expected as decrease in density of WDM would require increase in decay width owing to the fact that the observed X-ray flux is proportional to dark matter density times the decay width. This increase in decay width can be obtained either by lowering the mass of loop particles like $W_R$ or increasing the mixing angle $\theta_{\nu}$, as seen from the left panel plot of figure \ref{fig7a}.

\begin{figure}[!h]
\centering
\epsfig{file=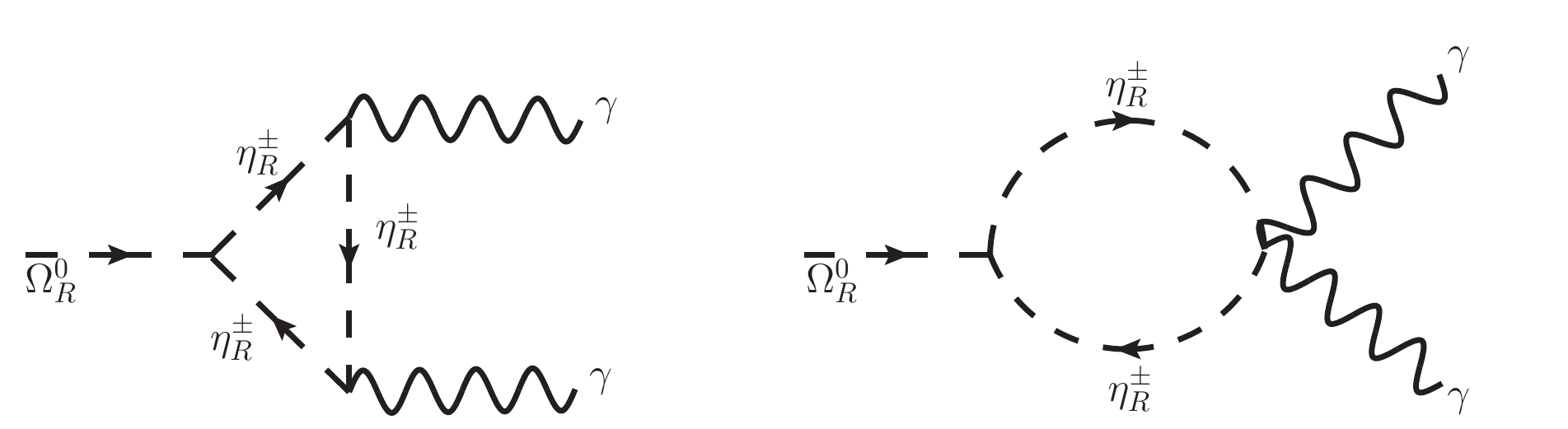,width=0.75\textwidth,clip=}
\caption{Scalar WDM decaying into two photons at one loop.}
\label{fig8}
\end{figure}

\begin{figure}[!h]
\centering
\epsfig{file=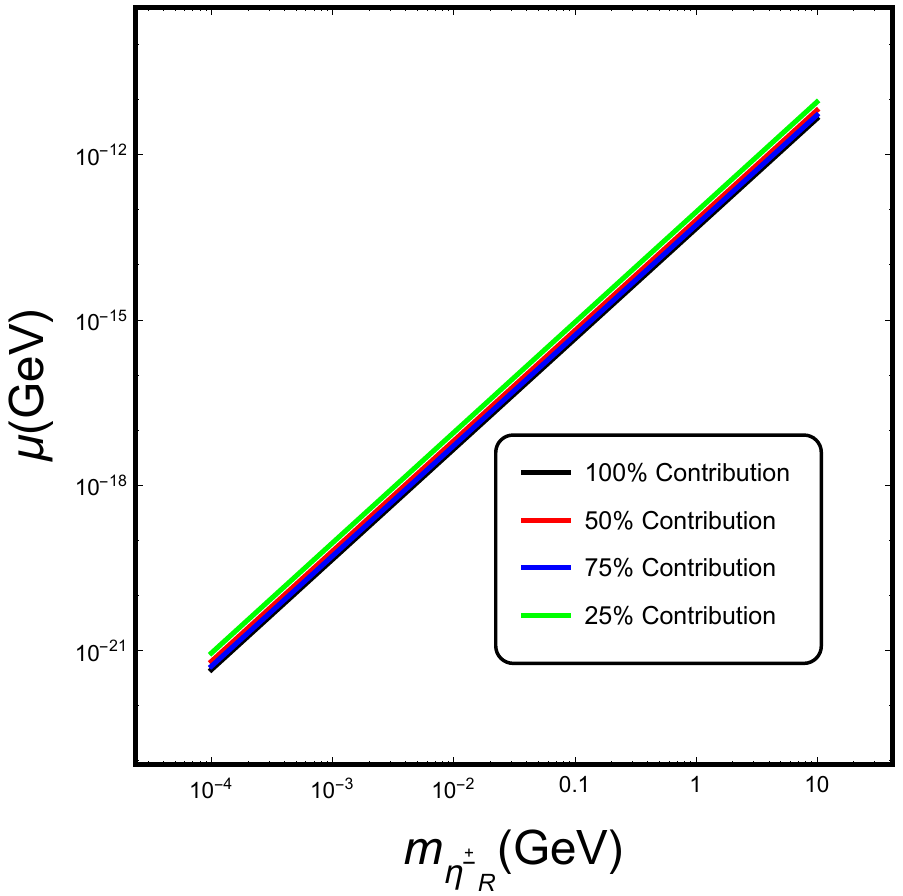,width=0.75\textwidth,clip=}
\caption{Parameter space giving rise to a long lived scalar WDM decaying into two photons at one loop.}
\label{fig8a}
\end{figure}

For scalar WDM on the other hand, such a decay can occur at one loop level through charged component of CDM doublet as seen from figure \ref{fig8}. We note that, such keV scalar DM decaying into two photons at loop level was discussed in the context of the 3.55 keV line by the authors of \cite{Babu:2014uoa}. The Decay of the neutral component of the triplet $\Omega_R$ to two photons is given as
\begin{equation}
\Gamma_{\Omega^0_R \rightarrow \gamma \gamma} = \frac{\mu^2 e^4}{16\pi m_{\Omega^0_R}}|\mathcal{I}|^2
\end{equation}
where 
\begin{align}
|\mathcal{I}|^2 &= \frac{1}{256\pi^4}\left(4|\mathcal{A}|^2 + \Re(\mathcal{A}^*\mathcal{B})\right) \\
\mathcal{I}^{\mu\nu} &= \frac{i}{16\pi^2}\left(\mathcal{A}g^{\mu \nu} + \frac{k^\mu_2 k^\nu_1}{m^2_{\Omega}}\mathcal{B}\right) \\
\mathcal{A} &= \frac{i}{16\pi^2}\left(1 + t\ln\left[\frac{2t-1+\sqrt{1-4t}}{2t}\right]^2\right) \\
\mathcal{B} &= -2\mathcal{A}
\end{align}
with $t = m^2_{\eta_R}/m^2_{\Omega^0_R}$. In the above expression for decay width, $\mu$ corresponds to the trilinear mass term involved in the coupling $\eta^{\dagger}_R \Omega_R \eta_R$. We fix the mass of $\Omega^0_R$ at 7.1 keV and vary the other two free parameters, namely $\mu, m_{\eta_R}$ from the requirement of the lifetime of $\Omega^0_R$ to be around $10^{28}$ s. The resulting parameter space is shown in figure \ref{fig8a}, where the effect of changing the relative abundance of WDM on allowed parameter space is marginal compared to the fermionic WDM case. But the overall upward shift with decrease in relative abundance is expected as increase in the parameter $\mu$ increases the decay width. Such a small trilinear coupling $\mu$ between $\Omega_R$ and $\eta_R$ by invoking the presence of additional symmetries.

\section{Conclusion}
\label{sec7}
We have studied a class of left right symmetric models where the dark matter sector can consist of a keV scale warm component and a GeV-TeV scale cold component. Since both the DM components have gauge interactions, they can be thermally produced in the early Universe. While the cold component's relic can be produced through the usual WIMP freeze-out mechanism, the warm component typically gets overproduced. This requires late time entropy dilution to bring the overproduced warm dark matter relic density to the observed or under-abundant regime. This requires a mother particle with a relatively long lifetime of 1 s or less, which we consider to be one of the right handed neutrinos.

The minimal LRSM can be extended to accommodate cold DM component in a straightforward manner, without the need of any additional discrete symmetries. If the lightest right handed neutrino is the keV WDM candidate, it is required to have tiny mixing with the left handed neutrinos to acquire a long lifetime, required for a DM candidate. This also helps in entropy dilution required to dilute the overproduced keV DM, by generating a long lifetime $(\sim 1 \; \text{s})$ of the decaying particle responsible for generating entropy. Such small Yukawa couplings can still be consistent with correct neutrino masses due to additional type II seesaw contribution.

We study three different cold dark matter candidates namely, a left scalar doublet, a right scalar doublet and a right fermion triplet and find the parameter space that can give rise to total as well as sub-dominant DM density. Since left fermion triplet relic abundance depends only on its mass (a fixed value around 3 TeV \cite{Ma:2008cu}), we do not pursue it in this work. For other CDM candidates, we constrain the right sector gauge boson mass as well as DM-Higgs portal couplings from the relic abundance criteria. We then calculate the thermal relic abundance of keV fermion DM (lightest right handed neutrino), and the required entropy dilution due to the decay of the next to lightest right handed neutrino. This requires the next to lightest right handed neutrino to be in sub-GeV regime and right handed gauge boson masses around TeV scale. This agrees with earlier works on WDM in LRSM \cite{Nemevsek:2012cd, Bezrukov:2009th}. We also do the analysis for a keV scalar DM for the sake of completeness, though the generic conclusions obtained in fermion WDM scenario do not change significantly in scalar WDM case, if both the masses are in keV regime.

We finally show the most interesting aspect of such a scenario that is, the indirect detection prospects of such mixed DM scenario. While the CDM component can have interesting signatures at gamma ray telescopes like the Fermi-LAT, the keV DM can give rise to monochromatic X-ray line if it decays on cosmological scales at radiative level. We constrain the parameter space for cold dark matter from the the latest gamma ray bounds from experiments like the Fermi-LAT. We also find the relevant parameter space such that either a fermion or a scalar WDM with 7.1 keV mass can give rise to a monochromatic 3.55 keV X-ray line, as claimed to be present in the XMM-Newton telescope data.

It should be noted that WDM with mass at keV scale can face constraints from structure formation data. As noted in \cite{Boyarsky:2008xj}, Lyman-$\alpha$ bounds restrict the keV fermion mass to be above 8 keV if it is non-resonantly produced and contributes $100\%$ to the total DM abundance. However, for less than $60\%$ contribution to total DM, such strict mass bounds do not apply. As shown in another recent work \cite{Kamada:2016vsc} which studies constraints on mixed DM from anomalous strong systems, such mixed DM scenario with WDM component less than $47\%$ of total DM abundance is compatible with small scale structure formation as well as 3.55 keV X-ray line data, at $95\%$ CL. Another stringent limit on WDM mass was derived recently from the abundance of ultra faint galaxies in the Hubble frontier fields \cite{ Menci:2016eui}. The authors derived the bounds based on different production mechanisms. While the bound on non-resonantly produced WDM mass is similar to the above mentioned references, the corresponding bound on a thermal relic WDM mass is weaker $M_{\rm WDM} > 2.9$ keV which agrees with the mass limit considered in our work as well as earlier works on thermal relic WDM \cite{Nemevsek:2012cd, Bezrukov:2009th}.

 It should be noted that, here we present the idea of such multi-component keV-TeV DM in a very simplified way that allows us to calculate their thermal abundance separately. This is justified for the type of CDM and WDM candidates chosen so that CDM freezes out earlier, followed by the freeze-out of WDM and then we have entropy release just before the BBN epoch. Also, the CDM in the model do not have efficient annihilation channels to the WDM candidates. In a general setup, one has to solve the coupled Boltzmann equations for the two candidates along with the decaying particle responsible for entropy release for more accurate results. We leave such a detailed study to future works.

\acknowledgments
DB acknowledges the support from IIT Guwahati start-up grant (reference number: xPHYSUGIITG01152xxDB001) and Associateship Programme of IUCAA, Pune. 

\appendix
\section{Scalar Potential of the Model}
\label{appen1}
The scalar potential for the minimal LRSM is 
\begin{equation}
 V(\Phi,\Delta_L,\Delta_R) = V_{\mu} + V_{\Phi} + V_{\Delta} + V_{\Phi\Delta} + V_{\Phi\Delta_L\Delta_R},
\end{equation}
where the bilinear terms in Higgs fields are
\begin{eqnarray}
 V_{\mu} &=& -\mu_1^2 \Tr{\Phi^\dagger\Phi}
 - \mu_2^2\Tr{\Phi^\dagger\tilde{\Phi} + \tilde{\Phi}^\dagger\Phi}
 - \mu_3^2\Tr{\Delta_L^\dagger\Delta_L + \Delta_R^\dagger\Delta_R}.
 \end{eqnarray}
The self-interaction terms of $\Phi$ are: 
\begin{align}
 V_{\Phi} = ~
  & \lambda_1\left[\Tr{\Phi^\dagger\Phi}\right]^2 +
  \lambda_2\left[\Tr{\Phi^\dagger\tilde{\Phi}}\right]^2 + \lambda_2\left[\Tr{\tilde{\Phi}^\dagger\Phi}\right]^2 
  \nonumber \\   
  & + \lambda_3\Tr{\Phi^\dagger\tilde{\Phi}}\Tr{\tilde{\Phi}^\dagger\Phi} + 
  \lambda_4\Tr{\Phi^\dagger\Phi}\Tr{\Phi^\dagger\tilde{\Phi} + \tilde{\Phi}^\dagger\Phi}.
  \label{appeneq1}
\end{align}
and the $\Delta_{L,R}$ self- and cross-couplings are as follows:
\begin{align}
 V_{\Delta} = ~ 
 & \rho_1\left(\left[\Tr{\Delta_L^\dagger\Delta_L}\right]^2 + 
     \left[\Tr{\Delta_R^\dagger\Delta_R}\right]^2\right) +
 \rho_3\Tr{\Delta_L^\dagger\Delta_L}\Tr{\Delta_R^\dagger\Delta_R}
 \nonumber \\ 
 & + \rho_2 \left(\Tr{\Delta_L\Delta_L}\Tr{\Delta_L^\dagger\Delta_L^\dagger} + \Tr{\Delta_R\Delta_R}\Tr{\Delta_R^\dagger\Delta_R^\dagger} \right)
\nonumber  \\ 
 & + \rho_4\left(\Tr{\Delta_L\Delta_L}\Tr{\Delta_R^\dagger\Delta_R^\dagger} + \Tr{\Delta_L^\dagger\Delta_L^\dagger}\Tr{\Delta_R\Delta_R}\right).
   \label{appeneq2}
\end{align}
In addition, there are also  $\Phi-\Delta_L$ and $\Phi-\Delta_R$ interactions present in the model, 
\begin{align}
 V_{\Phi\Delta} = ~
 & \alpha_1\Tr{\Phi^\dagger\Phi}\Tr{\Delta_L^\dagger\Delta_L + \Delta_R^\dagger\Delta_R} +
 \alpha_3 \Tr{\Phi\Phi^\dagger\Delta_L\Delta_L^\dagger + \Phi^\dagger\Phi\Delta_R\Delta_R^\dagger} 
  \nonumber \\ 
 & + \left\{
 \alpha_2 e^{i\delta_2}\Tr{\Phi^\dagger\tilde{\Phi}}\Tr{\Delta_L^\dagger\Delta_L} + 
 \alpha_2 e^{i\delta_2}\Tr{\tilde{\Phi}^\dagger\Phi}\Tr{\Delta_R^\dagger\Delta_R} + \text{H.c.}\right\}
  \label{appeneq3}
\end{align}
with $\delta_2 =0$ making CP conservation explicit, and the $\Phi-\Delta_L-\Delta_R$ couplings are
\begin{align}
 V_{\Phi\Delta_L\Delta_R} = ~
 & \beta_1\Tr{\Phi^\dagger\Delta_L^\dagger\Phi\Delta_R + \Delta_R^\dagger\Phi^\dagger\Delta_L\Phi} +
 \beta_2\Tr{\Phi^\dagger\Delta_L^\dagger\tilde{\Phi}\Delta_R + \Delta_R^\dagger\tilde{\Phi}^\dagger\Delta_L\Phi}
\nonumber \\ 
& + \beta_3\Tr{\tilde{\Phi}^\dagger\Delta_L^\dagger\Phi\Delta_R + \Delta_R^\dagger\Phi^\dagger\Delta_L\tilde{\Phi}}.
\end{align}

\bibliographystyle{apsrev}
\bibliography{ref.bib}

\end{document}